\documentclass[twocolumn,aps,prb,nofootinbib,superscriptaddress]{revtex4-2}
\usepackage{dcolumn}
\usepackage{graphicx}
\usepackage{bm}
\usepackage[linktocpage=true,colorlinks=true,urlcolor=blue,linkcolor=red,citecolor = blue]{hyperref}
\usepackage{amsfonts}
\usepackage{amsmath}
\usepackage{amssymb}
\usepackage{physics}
\usepackage[dvipsnames]{xcolor}

\usepackage[T1]{fontenc}
\graphicspath{ {Figures/} }

\begin{document}

\title{Spin-orbit enhancement in Si/SiGe heterostructures with oscillating Ge concentration}

\author{Benjamin D. Woods}
\affiliation{Department of Physics, University of Wisconsin-Madison, Madison, Wisconsin 53706, USA}
\author{M. A. Eriksson}
\affiliation{Department of Physics, University of Wisconsin-Madison, Madison, Wisconsin 53706, USA}
\author{Robert Joynt}
\affiliation{Department of Physics, University of Wisconsin-Madison, Madison, Wisconsin 53706, USA}
\author{Mark Friesen}
\affiliation{Department of Physics, University of Wisconsin-Madison, Madison, Wisconsin 53706, USA}

\begin{abstract}
We show that Ge concentration oscillations within the quantum well region of a Si/SiGe heterostructure can significantly enhance the spin-orbit coupling of the low-energy conduction-band valleys. Specifically, we find that for Ge oscillation wavelengths near $\lambda = 1.57~\text{nm}$ with an average Ge concentration of $\bar{n}_{\text{Ge}} = 5\%$ in the quantum well region, a Dresselhaus spin-orbit coupling is induced, at all physically relevant electric field strengths, which is over an order of magnitude larger than what is found in conventional Si/SiGe heterostructures without Ge concentration oscillations. 
This enhancement is caused by the Ge concentration oscillations producing wave-function satellite peaks a distance $2 \pi/\lambda$ away in momentum space from each valley, which then couple to the opposite valley through Dresselhaus spin-orbit coupling. Our results indicate that the enhanced spin-orbit coupling can enable fast spin manipulation within Si quantum dots using electric dipole spin resonance in the absence of micromagnets. Indeed, our calculations yield a Rabi frequency $\Omega_{\text{Rabi}}/B > 500~\text{MHz/T}$ near the optimal Ge oscillation wavelength $\lambda = 1.57~\text{nm}$. 
\end{abstract}

\maketitle

\section{Introduction}

Following the seminal work of Loss and DiVincenzo \cite{Loss1998}, quantum dots in semiconductors have emerged as a leading candidate platform for quantum computation \cite{Hanson2007,Burkard2021,Chatterjee2021,Zhang2019}. Gate-defined quantum dots in silicon \cite{Zwanenburg2013,Gyure2021} are particularly attractive due to their compatibility with the microelectronics fabrication industry. Moreover, in contrast with GaAs, for which coherence times are limited by unavoidable hyperfine interactions with nuclear spins \cite{Petta2005}, isotropic enrichment dramatically suppresses these interactions in Si, enabling long coherence times \cite{Assali2011}.

While recent progress in Si quantum dots has been quite promising, many of the leading qubit architectures rely on synthetic spin-orbit coupling arising from micromagnets \cite{PioroLadriere2008,Xue2022,Noiri2022,Mills2021}, leading to challenges for scaling up to systems with many dots. An alternative approach is to use intrinsic spin-orbit coupling for qubit manipulation, for example, through the electric dipole spin resonance (EDSR) mechanism~\cite{Golovach2006,Rashba2008}. While this possibility has been considered for Ge and Si hole-spin qubits, where the degeneracy of the $p$-orbital-dominated valence band leads to strong spin-orbit coupling \cite{Winkler2003, Terrazos2021}, the weak spin-orbit coupling of the Si conduction band appears unfavorable for electron-spin qubits. 

\begin{figure}[t]
\begin{center}
\includegraphics[width=0.48\textwidth]{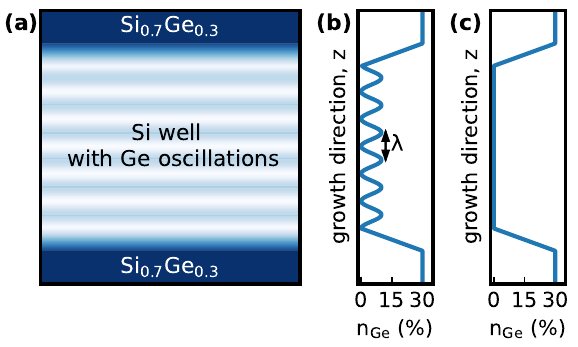}
\end{center}
\vspace{-0.5cm}
\caption{(a) Schematic of the Si/SiGe heterostructure considered here, consisting of a Si-dominated quantum well sandwiched between Si\textsubscript{0.7}Ge\textsubscript{0.3} barrier regions. Note that the growth direction is along the $\left[001\right]$ crystallographic axis. (b) Ge concentration profile along the growth ($z$) direction of a wiggle well. Ge concentration oscillations of wavelength $\lambda$ inside of the quantum well region lead to spin-orbit coupling enhancement for a proper choice of $\lambda$. (c) Ge concentration profile of a ``conventional'' Si/SiGe quantum well, for comparison.}
\label{FIG1}
\vspace{-1mm}
\end{figure}

In this work, we show how the spin-orbit coupling 
in Si/SiGe quantum well heterostructures can be enhanced by more than an order of magnitude by incorporating Ge concentration oscillations \textit{inside} the quantum well, leading to the possibility of exploiting intrinsic spin-orbit coupling in Si quantum dots for fast gate operations. Figure~\ref{FIG1}(a) shows a schematic of the system which consists of a Si-dominated quantum well region sandwiched between $\text{Si}_{0.7}\text{Ge}_{0.3}$ barrier regions, where the growth direction is taken along the $\left[001\right]$ crystallographic axis. In contrast to ``conventional'' Si/SiGe quantum wells, the quantum well region contains a small amount of Ge with concentration oscillations of wavelength $\lambda$, as shown in Fig.~\ref{FIG1}(b). For comparison, Fig.~\ref{FIG1}(c) shows the Ge concentration profile of a conventional Si/SiGe quantum well. Previous works  \cite{McJunkin2021,Feng2022} have studied such a structure, which has been named the wiggle well, and found that the periodic Ge concentration leads to an enhancement of the valley splitting. Here, we develop a theory of spin-orbit coupling within such structures and show that the periodic nature of the device, along with the underlying diamond crystal structure and degeneracy of the Si $z$ valleys, also gives rise to an enhancement in spin-orbit coupling. Importantly, we find that the wavelength $\lambda$ must satisfy a resonance condition to give rise to this spin-orbit coupling enhancement. As discussed in detail in Sec. \ref{Mechanism}, this involves a two-step process that can be summarized as follows. First, the periodic potential produced by the Ge concentration oscillations produces wave-function satellites a distance $2\pi/\lambda$ away in momentum space from each valley. Then, a satellite of a given valley couples strongly to the opposite valley through Dresselhaus spin-orbit coupling, provided that the satellite-valley separation distance in momentum space is $4\pi/a$, corresponding to the condition $\lambda = 1.57~\text{nm}$.


From the outset, it is important to remark that the spin-orbit coupling introduced by the Ge concentration oscillations is fundamentally distinct from the spin-orbit coupling of conventional Si/SiGe quantum wells. For a given subband of a conventional Si/SiGe quantum well immersed in a vertical electric field, the $C_{2v}$ point group symmetry of the system allows for both Rashba and ``Dresselhaus-type'' linear-$\mathbf{k}_\parallel$ terms of the form \cite{Nestoklon2008,Prada2011},
\begin{equation}
    H_\text{SO} = \alpha \left(k_y\bar{\sigma}_x - k_x \bar{\sigma}_y\right)
    + \beta \left(k_x \bar{\sigma}_x - k_y \bar{\sigma}_y\right),
\end{equation}
where $\bar{\sigma}_j$ are the Pauli matrices acting in (pseudo)spin space and $\alpha$ and $\beta$ are the Rashba and Dresselhaus coefficients, respectively, of the subband. The presence of Rasbha spin-orbit coupling is unsurprising due to the structural asymmetry provided by the electric field \cite{Bychkov1984}, while the presence of the Dresselhaus-type term is initially surprising since the diamond lattice of Si/SiGe quantum wells possesses bulk inversion \textit{symmetry}~\cite{Dresselhaus1955}. However, these systems still support a Dresselhaus-type term of the same form $\beta(k_x \bar{\sigma}_x - k_y \bar{\sigma}_y)$, due to the broken inversion symmetry caused by the quantum well interfaces~\cite{Ivchenko1996,Nestoklon2008,Prada2011,Vervoort1997}. This is in stark contrast to the true Dresselhaus spin-orbit coupling in III-V semiconductors, where the asymmetry of the anion and cation in the unit cell leads to bulk inversion asymmetry~\cite{Dresselhaus1955}. Importantly, we find in Sec.~\ref{WiggleWellResults} that the spin-orbit coupling of the wiggle well 
does not rely upon the presence of an interface. Rather, it is an intrinsic property of a \textit{bulk} system with Ge concentration oscillations. In this sense, the spin-orbit coupling investigated here is more akin to the true Dresselhaus spin-orbit coupling of III-V semiconductors than the Dresselhaus-type spin-orbit coupling of conventional Si/SiGe quantum wells brought about by interfaces. Indeed, the only requirement for linear-$\mathbf{k}_\parallel$ Dresselhaus spin-orbit coupling in a wiggle well with an appropriate $\lambda$ is confinement in the growth direction (even symmetric confinement), to allow for the formation of subbands.
For simplicity in the remainder of this work, we simply refer to this form of spin-orbit coupling as Dresselhaus.

The rest of this paper is organized as follows. In Sec.~\ref{Model} we describe our model used to study the quantum well heterostructure. 
Section \ref{Results} then presents our numerical results for the spin-orbit coefficients. This also includes the calculation of the EDSR Rabi frequency and studies the impact of alloy disorder on the spin-orbit coefficients. In Sec.~\ref{Mechanism} we provide an extensive explanation of the mechanism behind the spin-orbit coupling enhancement. 
Finally, we conclude in Sec.~\ref{Conclusion}.

\section{Model} \label{Model}
In this section, we outline the model used to study our Si/SiGe heterostructure along with the methods used to calculate the spin-orbit coefficients. In Sec.~\ref{AlloyModel}, we describe the tight binding model used to model generic SiGe alloy systems. Next, in Sec.~\ref{VCModel} we employ a virtual crystal approximation to impart translation invariance in the plane of the quantum well, allowing us to reduce the problem to an effective one-dimensional (1D) Hamiltonian parametrized by in-plane momentum $\mathbf{k}_\parallel$. In Sec.~\ref{ExpansionSec}, we expand the model around $\mathbf{k}_\parallel = 0$ to separate out the Hamiltonian components that give rise to Rashba and Dresselhaus spin-orbit coupling, respectively, and we explain the important differences between the two components. Finally, in Sec.~\ref{SubbandModel}, we transform the Hamiltonian into the subband basis, which allows us to obtain expressions for the Rashba and Dresselhaus spin-orbit coefficients in each subband.

\subsection{Model of SiGe alloys} \label{AlloyModel}
To study the spin-orbit physics of our system we use the empirical tight-binding method \cite{Slater1954}, where the electronic wave function is written as a linear combination of atomic orbitals:
\begin{equation}
    \ket{\psi} =
    \sum_{n,j,\nu,\sigma}
    \ket{nj\nu\sigma} \psi_{nj\nu\sigma}.
\end{equation}
Here, $\bra{\boldsymbol{r}}\ket{nj\nu \sigma} = \phi_\nu\left(\boldsymbol{r} - \boldsymbol{R}_{n,j}\right)\ket{\sigma}$ is an atomic orbital centered at position $\boldsymbol{R}_{n,j}$, corresponding to atom $j$ of atomic layer $n$ along the growth direction $[001]$, $\nu$ is a spatial orbital index, and $\ket{\sigma}$ is a two component spinor with $\sigma = \uparrow,\downarrow$ indicating the spin of the orbital. We use an sp\textsuperscript{3}d\textsuperscript{5}s\textsuperscript{*} basis set with 20 orbitals per atom, onsite spin-orbit coupling, nearest-neighbor hopping, and strain. Note that nearest-neighbor sp\textsuperscript{3}d\textsuperscript{5}s\textsuperscript{*} tight-binding models are well established for accurately describing the electronic structure of semiconductor materials over a wide energy range \cite{Jancu1998}.   Explicitly, $\nu$ is a spatial orbital index from the set including $s$, $s^*$, $p_i$ ($i = x,y,z$), and $d_i$ ($i = xy,yz,zx,z^2,x^2 - y^2$) orbitals, which are meant to model the outer-shell orbitals of individual Si and Ge atoms that participate in chemical bonding. Additionally, these orbitals possess certain spatial symmetries that, combined with the diamond crystal structure of the SiGe alloy, dictate the forms of the nearest neighbor couplings, as first explained in the work of Slater and Koster \cite{Slater1954}. The free parameters of the tight-binding model (including onsite orbital energies, nearest-neighbor hopping energies, strain parameters, etc.) are then chosen such that the band structure of the system agrees as well as possible with experimental and/or \textit{ab initio} data. In this work, we use the tight binding model and parameters of Ref. \cite{Niquet2009}, which allows for the modeling of strained, random SiGe alloys with any Ge concentration profile. 

The Hamiltonian of an arbitrary SiGe alloy takes the form,
\begin{multline}
    H^{mi,nj}_{\mu \sigma, \nu \sigma} = \delta_{m i}^{n j} 
    \left[
    \delta_{\mu \sigma}^{\nu \sigma^\prime}
    \left(\varepsilon_{\nu}^{(n j)} + V_n\right) +  \delta_{\sigma}^{\sigma^\prime} C_{\mu\nu }^{(n j)}
    + S_{\mu \sigma, \nu \sigma^\prime}^{(n j)} \right]
    \\
    + \delta_{\sigma}^{\sigma^\prime}\left(\delta_{m}^{n+1} T^{(n)}_{i \mu,j \nu}
    +  \delta_{m}^{n-1} T^{(m)\dagger}_{i \mu,j \nu}\right) ,
    \label{HamRS1}
\end{multline}
where $H^{mi,nj}_{\mu \sigma, \nu \sigma^\prime} = \mel{mi\mu\sigma}{H}{n j \nu\sigma^\prime}$ and $\delta$ equals $1$ if its subscripts match its superscripts and $0$ otherwise. The first line in Eq.~(\ref{HamRS1}) contains \textit{intra-atomic} terms, where $\varepsilon_{\nu}^{(n j)}$ is the onsite energy of orbital $\nu$, for atom $j$ in atomic layer $n$, $V_n$ is the potential energy due to the vertical electric field, $C_{\mu\nu }^{(n j)}$ accounts for onsite energy shifts and couplings caused by strain, and $S_{\mu \sigma, \nu \sigma^\prime}^{(n j)}$ accounts for spin-orbit coupling. The matrix $C_{\mu\nu }^{(n j)}$ is determined by the deformation of the lattice due to strain as detailed in Ref.~\cite{Niquet2009} and arises from changes in the onsite potential of the atom due to the displacement of its neighbors. In addition, spin-orbit coupling $S^{(nj)}$ is an intra-atomic coupling between $p$ orbitals~\cite{Chadi1977} and is the only term in Eq.~(\ref{HamRS1}) that does not conserve spin $\sigma$. (See Appendix \ref{PSBTM} for the explicit form of $S^{(nj)}$.) 
Note that the superscripts $(nj)$, which index the atoms, are needed here because the intra-atomic terms depend on whether an atom is Si or Ge, as well as the local strain environment. 
The second line in Eq.~(\ref{HamRS1}) contains \textit{inter-atomic} terms describing the hopping between atoms on adjacent atomic layers, where $T^{(n)}$ is the hopping matrix from atomic layer $n$ to atomic layer $n+1$. Nearly all elements of $T^{(n)}$ are zero, with non-zero hoppings occurring only between nearest-neighbor atoms. A non-zero hopping matrix element $T^{(n)}_{i \mu,j \nu}$ then depends on three things: (1) the orbital indices $\mu$ and $\nu$, (2) the types of atoms involved, and (3) the direction and magnitude of the vector $\mathbf{R}_{n+1,i} - \mathbf{R}_{n,j}$ connecting the atoms. We then use the Slater-Koster table in Ref. \cite{Slater1954} along with the parameters of Ref.~\cite{Niquet2009} to calculate $T^{(n)}_{i \mu,j \nu}$. We note that strain affects the hopping elements 
by altering the direction and length of the nearest-neighbor vectors (i.e., the crystalline bonds)~\cite{Ren1982,Niquet2009}.

In this work, we let the Ge concentration vary between layers, as shown in Figs.~\ref{FIG1}(b) and \ref{FIG1}(c), but assume it to be uniform within a given layer. Note that the large difference in Ge concentration between the barrier and well regions results in a large conduction band offset that traps electrons inside the quantum well. This occurs naturally in the tight binding model of Eq.~(\ref{HamRS1}) because $\varepsilon^{(nj)}, C^{(nj)}$, and $T^{(n)}_{ij}$ are different for Si and Ge atoms. Finally, 
we point the reader to Appendix~\ref{perpLatConstant} for a description of the lattice constant dependence on strain.

\subsection{Virtual crystal approximation and pseudospin transformation} \label{VCModel}
\begin{figure}[t]
\begin{center}
\includegraphics[width=0.48\textwidth]{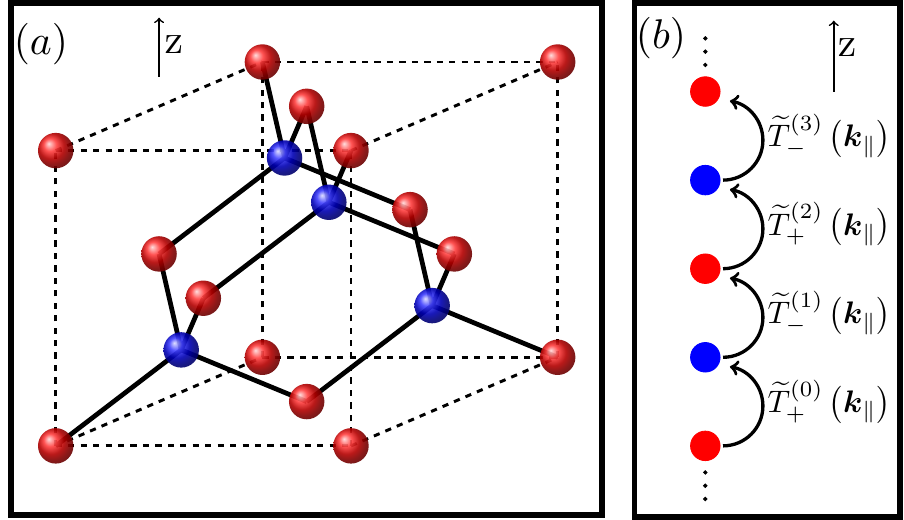}
\end{center}
\vspace{-0.5cm}
\caption{(a) Diamond crystal structure of silicon. The dashed lines outline the conventional unit cell of the face centered cubic lattice. Both red and blue atoms are silicon but belong to different sublattices. Notice that the vectors connecting an atom to its four nearest neighbors are fundamentally different for the red and blue atoms, giving rise to the alternating hopping structure shown in (b). (b) Effective 1D tight-binding chain, with hopping matrix terms alternating between $\widetilde{T}_+^{(n)}(\boldsymbol{k}_\parallel)$ and $\widetilde{T}_-^{(n)}(\boldsymbol{k}_\parallel)$. 
Note that each site has 20 orbitals, and only the forward hopping terms are shown. Onsite and backward hopping terms are not shown. For a SiGe alloy in the virtual crystal approximation, the two-sublattice structure is retained, but the atoms are replaced by virtual atoms, with averaged properties consistent with the Ge concentration of a given layer.}
\label{FIG2}
\vspace{-1mm}
\end{figure}

While the Hamiltonian in Eq. (\ref{HamRS1}) provides an accurate description of SiGe alloys, it lacks translation invariance when alloy disorder is present. This makes the model computationally expensive to solve, and it obscures the physics of the spin-orbit enhancement coming from the \textit{averaged} effects of the inhomogeneous Ge concentration profile. We therefore employ a virtual crystal approximation where the Hamiltonian matrix elements are replaced by their value averaged over all alloy realizations. Specifically, we define a virtual crystal Hamiltonian $H_\text{VC}$ with elements $(H_\text{VC})^{mi,nj}_{\mu \sigma, \nu \sigma^\prime} = \left< H^{mi,nj}_{\mu \sigma, \nu \sigma^\prime} \right>$, where $\left<\dots\right>$ indicates an average over all possible alloy realizations. The Hamiltonian is then 
translation invariant within the plane of the quantum well. In addition, it is useful to move beyond the original orbital basis, where the spin is well-defined, to a \textit{pseudospin} basis defined by
\begin{equation}
    \ket{n j \bar{\nu} \bar{\sigma}} = 
    \sum_{\nu \sigma} 
    \ket{ n j \nu\sigma}  
    U^{(n)}_{\nu\sigma,\bar{\nu}\bar{\sigma}}, \label{eq:pspin}
\end{equation}
where $\bar{\nu}$ are the hybridized orbital states, $\bar{\sigma} = \Uparrow,\Downarrow$ are the pseudospins, and $U^{(n)}$ is the transformation matrix for layer $n$. Full details of this basis transformation can be found in Appendix~\ref{PSBTM}. Here, we mention three important features of Eq.~(\ref{eq:pspin}). First, the basis transformation diagonalizes the onsite spin-orbit coupling, making the spin-orbit physics more transparent.
Second, the pseudospin states represent linear combinations of orbitals including both spin $\uparrow$ and spin $\downarrow$. Third, the transformation matrix $U^{(n)}$ can be shown to satisfy
\begin{equation}
    U^{(n)} = 
    \begin{cases}
       U^{(0)}, & n \in \mathbb{Z}_{\text{even}}, \\
        U^{(1)}, & n \in \mathbb{Z}_{\text{odd}}.   
    \end{cases}
\end{equation}
This alternating structure for the transformation matrix is crucial to the results that follow, and results from the presence of two sublattices in the diamond crystal structure of Si, as shown in Fig.~\ref{FIG2}(a).

Making use of the virtual crystal approximation, we now convert our three-dimensional (3D) Hamiltonian into an effective 1D Hamiltonian. To begin, we note that the virtual crystal Hamiltonian takes the simplified form,
\begin{multline}
    \left(H_\text{VC}\right)^{mi,nj}_{\bar{\mu} \bar{\sigma}, \bar{\nu} \bar{\sigma}^\prime} = \delta_{mi \bar{\sigma}}^{n j \bar{\sigma}^\prime} 
    \left[
    \delta_{\bar{\mu}}^{\bar{\nu}}\left(\bar{\varepsilon}_{\bar{\nu}}^{(n)} + V_n\right) 
    + \bar{C}_{\bar{\mu}\bar{\nu}}^{(n)}
    \right] \\
    + \delta_{m}^{n+1} \bar{T}^{(n)}_{i \bar{\mu} \bar{\sigma},j \bar{\nu} \bar{\sigma}^\prime}
    +  \delta_{m}^{n-1} \bar{T}^{(m)\dagger}_{i \bar{\mu} \bar{\sigma},j \bar{\nu} \bar{\sigma}^\prime}
 , \label{HamRS}
\end{multline}
where $\left(H_\text{VC}\right)^{mi,nj}_{\bar{\mu} \bar{\sigma}, \bar{\nu} \bar{\sigma}^\prime} = \mel{mi\bar{\mu}\bar{\sigma}}{H_\text{VC}}{n j \bar{\nu}\bar{\sigma}^\prime}$, $\bar{\varepsilon}_{\bar{\nu}}^{(n)}$ is the onsite energy of pseudospin orbital $\bar{\nu}$, and  $\bar{C}^{(n)}$ and $\bar{T}^{(n)}$ are the onsite strain and hopping matrices, respectively, transformed into the pseudospin basis and averaged over alloy realizations. Note that $\bar{\varepsilon}^{(n)}$ includes contributions from diagonalizing the onsite spin-orbit coupling. Importantly, $\bar{\varepsilon}^{(n)}$ and $\bar{C}^{(n)}$ maintain a dependence on the layer index $n$ due to the non-uniform Ge concentration profile. 
In contrast, the dependence of intra-atomic terms on the \textit{intra}-layer atom index $j$ has vanished due to the virtual crystal approximation. Moreover, the translation invariance of the virtual crystal approximation implies that hopping matrix elements between any two layers only depend upon the relative position of atoms, i.e., $\bar{T}^{(n)}_{i\bar{\mu} \bar{\sigma},j\bar{\nu} \bar{\sigma}^\prime} = \bar{T}^{(n)}_{\bar{\mu} \bar{\sigma}, \bar{\nu}\bar{\sigma}^\prime}(\boldsymbol{R}_{n+1,i} - \boldsymbol{R}_{n,j})$. We therefore introduce in-plane momentum $\boldsymbol{k}_\parallel$ as a good quantum number and Fourier transform our Hamiltonian.
To do so, we define the basis state
\begin{equation}
   \ket{\boldsymbol{k}_\parallel n \bar{\nu} \bar{\sigma}}
    =
    \frac{1}{\sqrt{N_\parallel}}
    \sum_j e^{i \boldsymbol{k_\parallel} \cdot \boldsymbol{R}_{nj}} \ket{nj \bar{\nu}\bar{\sigma}} ,
\end{equation}
where $\boldsymbol{k}_\parallel = (k_x,k_y)$ and $ N_\parallel$ is the number of atoms within each layer. The Hamiltonian has matrix elements
\begin{multline}
   \widetilde{H}^{mn}_{\bar{\mu}\bar{\sigma},\bar{\nu}\bar{\sigma}^\prime}(\boldsymbol{k}_\parallel) =
   \delta_{m \bar{\sigma}}^{n \bar{\sigma}^\prime}
    \left[
    \delta_{\bar{\mu}}^{\bar{\nu}}\left(\bar{\varepsilon}_{\bar{\nu}}^{(n)} + V_n\right) 
    + \bar{C}_{\bar{\mu}\bar{\nu}}^{(n)}
    \right] \\
    + \delta_{m}^{n+1}  \widetilde{T}_{\bar{\mu}\bar{\sigma}, \bar{\nu}\bar{\sigma}^\prime}^{(n)}\left(\boldsymbol{k}_\parallel\right)
    + \delta_{m}^{n-1}  \widetilde{T}_{\bar{\mu}\bar{\sigma}, \bar{\nu}\bar{\sigma}^\prime}^{(m)\dagger} \left(\boldsymbol{k}_\parallel\right)
, \label{BlochHam}
\end{multline}
where $\widetilde{H}^{mn}_{\bar{\mu}\bar{\sigma},\bar{\nu}\bar{\sigma}^\prime}(\boldsymbol{k}_\parallel) = \mel{ \boldsymbol{k}_\parallel m\bar{\mu}\bar{\sigma}}{H_\text{VC}}{ \boldsymbol{k}_\parallel n\bar{\nu}\bar{\sigma}^\prime}$, and $\widetilde{T}_{\bar{\mu}\bar{\sigma}, \bar{\nu}\bar{\sigma}^\prime}^{(n)}\left(\boldsymbol{k}_\parallel\right)$ is the Fourier-transformed hopping matrix given by
\begin{equation}
    \widetilde{T}_{\bar{\mu}\bar{\sigma}, \bar{\nu}\bar{\sigma}^\prime}^{(n)}\left(\boldsymbol{k}_\parallel\right)
    = \sum_{l = 1}^2
    e^{-i \boldsymbol{k_\parallel} \cdot \boldsymbol{r}_l^{(n)}}
    \bar{T}_{\bar{\mu} \bar{\sigma}, \bar{\nu}\bar{\sigma}^\prime}^{(n)}(\boldsymbol{r}_l^{(n)}), \label{Ttilde}
\end{equation}
where $\boldsymbol{r}_l^{(n)}$ is a nearest-neighbor vector from a reference atom in layer $n$ to one of its nearest neighbors in layer $n+1$. For a diamond lattice, each atom has only has two such bonds, as indicated in Fig.~\ref{FIG2}(a). Note that the Hamiltonian matrix elements vanish between states with different momenta due to translational invariance. Hence, we obtain an effective 1D Hamiltonian as a function of $\boldsymbol{k}_\parallel$.

An important feature of the Fourier-transformed hopping matrix $\widetilde{T}^{(n)}(\boldsymbol{k}_\parallel)$ is that it depends on the layer index $n$ for two reasons. First, the inhomogeneous Ge concentration along the growth axis causes the hopping parameters to change slightly from layer to layer. Second, and more importantly, the diamond crystal structure is composed of two interleaving face-centered-cubic sublattices which each contribute an inequivalent atom to the primitive unit cell. This is illustrated in Fig.~\ref{FIG2}(a) where the atoms belonging to the two sublattices are colored red and blue, respectively. Indeed, the atoms for $n \in \mathbb{Z}_{\text{even}}$ and $n \in \mathbb{Z}_{\text{odd}}$ belong to sublattice 1 and 2, respectively, and have different nearest neighbor vectors.
It is therefore useful to define 
\begin{equation}
    \widetilde{T}^{(n)}\left(\boldsymbol{k}_\parallel\right) = 
    \begin{cases}
       \widetilde{T}_{+}^{(n)}\left(\boldsymbol{k}_\parallel\right), & n \in \mathbb{Z}_{\text{even}} \\
       \widetilde{T}_{-}^{(n)}\left(\boldsymbol{k}_\parallel\right), & n \in \mathbb{Z}_{\text{odd}}
    \end{cases} \label{TnEvenOdd}
\end{equation}
as the hopping matrices for the two sublattices. We stress that the dependence of $\widetilde{T}_{+}^{(n)}(\boldsymbol{k}_\parallel)$ and $\widetilde{T}_{-}^{(n)}(\boldsymbol{k}_\parallel)$ on the layer index $n$ is due to the inhomogeneous Ge concentration profile, 
 and that $\widetilde{T}^{(n)}_{+}(\boldsymbol{k}_\parallel)$ and $\widetilde{T}^{(n)}_{-}(\boldsymbol{k}_\parallel)$ differ due to the diamond crystal structure having two inequivalent atoms in its primitive unit cell.
We can therefore visualize the system, for any given $\boldsymbol{k}_\parallel$, as a 1D, multi-orbital tight binding chain, as shown in Fig.~\ref{FIG2}(b), where the hopping terms alternate in successive layers.


\subsection{Expansion around $\mathbf{k}_\parallel = 0$} \label{ExpansionSec}
Our goal is to understand the spin-orbit physics of low-energy conduction band states near the Fermi level. 
In strained Si/SiGe quantum wells, these derive from the two degenerate valleys near the $Z$-point of the strained Brillouin zone \cite{Schaffler1997}. Therefore, the low-energy states have small $|\mathbf{k}_\parallel|$, and we can understand the spin-orbit physics by expanding the Fourier-transformed hopping matrices $\widetilde{T}_{\pm}^{(n)}\left(\boldsymbol{k}_\parallel\right)$ to linear order.
We find that
\begin{equation}
    \widetilde{T}_{\pm}^{(n)}(\mathbf{k}_\parallel) = \widetilde{T}_0^{(n)} +  \widetilde{T}_R^{(n)}(\mathbf{k}_\parallel) 
    \pm \widetilde{T}_D^{(n)}(\mathbf{k}_\parallel) + \mathcal{O}(\mathbf{k}_\parallel^2), \label{Texpansion}
\end{equation}
where $\widetilde{T}_0^{(n)}$ is the hopping matrix for $\mathbf{k}_\parallel = 0$, and $\widetilde{T}_{R}^{(n)}$ and $\widetilde{T}_{D}^{(n)}$ contain the linear $\mathbf{k}_\parallel$ corrections. These hopping matrix components are found to be 
\begin{align}
    \widetilde{T}_0^{(n)} &= \Omega^{(n)} \bar{\sigma}_0, \label{To} \\
    \widetilde{T}_R^{(n)}(\mathbf{k}_\parallel)
    &=  \Phi^{(n)} 
    \left(k_y \bar{\sigma}_x - k_x \bar{\sigma}_y\right), \label{TRashba}\\
    \widetilde{T}_D^{(n)}(\mathbf{k}_\parallel)
    &= \Phi^{(n)}\left(k_x\bar{\sigma}_x - k_y \bar{\sigma}_y\right), \label{TD}
\end{align}
where $\Omega^{(n)}$ and $\Phi^{(n)}$ are real-valued $10 \times 10$ matrices, and $\bar{\sigma}_j$ are the Pauli matrices acting on pseudospin space, with $j=0,x,y,z$. 

There are several features to remark on in Eqs.~(\ref{To})-(\ref{TD}). First, the momentum-spin structure of $\widetilde{T}_R^{(n)}$ and $\widetilde{T}_D^{(n)}$ have the familiar forms of Rasbha and Dresselhaus spin-orbit coupling; hence, we apply the subscript labels $R$ and $D$. Second, the hopping matrices for the two sublattices, $\widetilde{T}_{+}^{(n)}$ and $\widetilde{T}_{-}^{(n)}$, differ by the sign in front of $\widetilde{T}_D^{(n)}$, while the sign of $\widetilde{T}_R^{(n)}$ is sublattice independent. As noted above, this alternating sign is a consequence of the two sublattices of the diamond crystal being inequivalent. 
As we shall show in Sec.~\ref{ExtPWB}, this alternating hopping structure of $\widetilde{T}_D^{(n)}$ is key to explaining the mechanism behind the enhanced spin-orbit coupling. In Appendix~\ref{SymArg}, we also provide a symmetry argument for why the system has this particular alternating hopping structure for the Rashba and Dresselhaus terms. Third, for the special case of $\mathbf{k}_\parallel = 0$, the two sublattices become equivalent (i.e., there is no even/odd structure due to the vanishing of $\widetilde{T}_R^{(n)}$ and $\widetilde{T}_D^{(n)}$ at $\mathbf{k}_\parallel = 0$). Furthermore, the pseudospin sectors are uncoupled and equivalent at $\mathbf{k}_\parallel = 0$, which implies that all eigenstates of our Hamiltonian are doubly degenerate at $\mathbf{k}_\parallel = 0$ as is expected since the system has time-reversal symmetry.
Fourth, the dependence of $\Omega^{(n)}$ and $\Phi^{(n)}$ on the layer index $n$ arises only from the inhomogeneous Ge concentration profile. However, note that neither matrix vanishes if the Ge concentration is uniform. For interested readers, we provide expressions for $\Omega^{(n)}$ and $\Phi^{(n)}$ in Appendix~\ref{OmegaPhi}, for the case of a pure Si structure. Finally, we note that the diagonal elements of $\Phi^{(n)}$ all vanish.

To linear order in $\mathbf{k}_\parallel$, the Hamiltonian then takes the compact form
\begin{multline}
    H = H_0^{(z)} \bar{\sigma}_0 + H_R^{(z)}\left(k_y \bar{\sigma}_x - k_x \bar{\sigma}_y\right)\\ + H_{D}^{(z)}\left(k_x \bar{\sigma}_x - k_y \bar{\sigma}_y\right) + \mathcal{O}(\mathbf{k}_\parallel^2)
, \label{Hk}
\end{multline}
where $H_0^{(z)}$ (which we call the subband Hamiltonian) describes the physics at $\mathbf{k}_\parallel = 0$, and $H_R^{(z)}$ and $H_D^{(z)}$ describe the linear $\mathbf{k}_\parallel$ perturbations arising from $\widetilde{T}_R^{(n)}$ and $\widetilde{T}_D^{(n)}$, respectively. These take the form
\begin{align}
    \begin{split}
    \mel{m \bar{\mu}}{H_0^{(z)}}{n \bar{\nu}} =& 
    ~\delta_{m}^{n}
    \left[
    \delta_{\bar{\mu}}^{\bar{\nu}}\left(\bar{\varepsilon}_{\bar{\nu}}^{(n)} + V_n\right) 
    + \bar{C}_{\bar{\mu}\bar{\nu}}^{(n)}
    \right] \\
    &+ \delta_{m}^{n+1} \Omega_{\bar{\mu} \bar{\nu}}^{(n)} 
    + \delta_{m}^{n-1} \Omega^{(m)T}_{\bar{\mu} \bar{\nu}} \end{split}, \label{Ho}\\
    \mel{m \bar{\mu}}{H_R^{(z)}}{n \bar{\nu}}=&
    \delta_{m}^{n+1} \Phi^{(n)}_{\bar{\mu} \bar{\nu}} 
    + \delta_{m}^{n-1} \Phi^{(m)T}_{\bar{\mu} \bar{\nu}}, \label{HR} \\
     \mel{m \bar{\mu}}{H_D^{(z)}}{n \bar{\nu}}=&
    (-1)^{n}  
    \left(\delta_{m}^{n+1} \Phi^{(n)}_{\bar{\mu} \bar{\nu}} 
    - \delta_{m}^{n-1} \Phi^{(m)T}_{\bar{\mu} \bar{\nu}}\right), \label{HD}
\end{align}
where the superscript $z$ indicates that only the orbital degrees of freedom in the $z$ direction (i.e., the growth direction) are acted upon. Hence, the momentum $\mathbf{k}_\parallel$ and pseudospin $\bar{\sigma}$ indices are both dropped in Eqs.~(\ref{Ho})-(\ref{HD}). Also note that the alternating $\pm$ factor in front of $\widetilde{T}_D^{(n)}$ in Eq.~(\ref{Texpansion}) is reflected in the $(-1)^{n}$ factor in Eq.~(\ref{HD}). 

\subsection{Transformation to the subband basis} \label{SubbandModel}
The largest term in Hamiltonian (\ref{Hk}) (by far) is the subband Hamiltonian $H_0^{(z)}$. The eigenstates of $H_0^{(z)}$ are referred to as the orbital subbands of the quantum well, including two distinct valley states per subband. 
The subband and valley states, in turn, serve as a natural basis for representing the Hamiltonian, since the lateral confinement associated with a quantum dot barely perturbs this subband designation, although disorder may cause hybridization of the valley states. It is therefore the properties of the individual subbands that largely determine the properties of quantum dot states, including their spin-orbit behavior. 

To perform a subband basis transformation, we define $\ket{\varphi_\ell}$ as the $\ell^\text{th}$ eigenstate of $H_0^{({z})}$ with energy $E_{\ell}$. 
(Here, for convenience, we include both subband and valley states in the set $\{\ell\}$.)
Generically we can write
\begin{equation}
    \ket{\varphi_\ell} = \sum_{n \bar{\nu}} 
    \ket{n \bar{\nu}} Q_{n \bar{\nu}, \ell},
\end{equation}
where $Q$ is an orthogonal matrix, defined such that
\begin{equation}
    H_0^{({z})}  \ket{\varphi_\ell} = E_\ell \ket{\varphi_\ell} 
\end{equation}
for each $\ell$. 
Using these eigenstates as a basis, the Hamiltonian can then be expressed as
\begin{multline}
    H = \bar{\Lambda} \sigma_0 + \bar{\alpha}\left(k_y \bar{\sigma}_x - k_x \bar{\sigma}_y\right) \\ 
    + \bar{\beta} \left(k_x \bar{\sigma}_x - k_y \bar{\sigma}_y\right)
    + \mathcal{O}(\mathbf{k}_\parallel^2) 
, \label{Hp}
\end{multline}
where $\bar{\Lambda}$, $\bar{\alpha}$, and $\bar{\beta}$ are real-symmetric matrices acting in subband space with elements
\begin{align}
    \bar{\Lambda}_{\ell \ell^\prime} =& \delta_{\ell\ell^\prime} E_{\ell}, \label{LambdaElems} \\
    \bar{\alpha}_{\ell \ell^\prime} =& \mel{\varphi_\ell}{H_R^{(z)}}{\varphi_{\ell^\prime}}, \label{alphaElems} \\
    \bar{\beta}_{\ell \ell^\prime} =& \mel{\varphi_\ell}{H_D^{(z)}}{\varphi_{\ell^\prime}}. \label{betaElems}
\end{align}
The matrix elements $\bar{\alpha}_{\ell \ell^\prime}$ and $\bar{\beta}_{\ell \ell^\prime}$ are referred to as the Rashba and Dresselhaus spin-orbit coupling coefficients, respectively \cite{Winkler2003}. The diagonal elements are of particular importance since they determine the linear dispersion of a given subband near $\mathbf{k}_\parallel=0$. Indeed, the diagonal elements $\bar{\alpha}_{\ell \ell}$ and $\bar{\beta}_{\ell \ell}$ themselves are often referred to in the literature as the Rasbha and Dresselhaus spin-orbit coupling coefficients, respectively, and are typically denoted simply as $\alpha$ and $\beta$. Furthermore, we focus on the diagonal elements of the ground ($\ell = 0$) and excited ($\ell = 1$) valley states corresponding to the lowest orbital subband, which we henceforth refer to as simply the ground and excited valley states. These represent the lowest-energy conduction subbands, which are nearly degenerate due to the wide separation of the two degenerate $z$ valleys within the Brillouin zone of Si~\cite{Friesen2007}, thus playing a dominating role in the physics of Si spin qubits. In some cases, we may also be interested in the spin-orbit coupling between the ground and excited valleys, often referred to as spin-valley coupling, 
since the valley states are much closer in energy than the orbitally excited subbands. We note that confinement in the growth direction is a crucial ingredient for obtaining nonzero values of $\bar{\alpha}_{\ell \ell}$ and $\bar{\beta}_{\ell \ell}$. 
While this latter fact is not obvious from the structure of the Hamiltonian, 
it can be shown to be true, using the fact that $\Phi^{(n)}$ has vanishing diagonal elements and the structure of the $\Omega^{(n)}$ and $\Phi^{(n)}$ matrices described in Appendix~\ref{OmegaPhi}.
Finally, we also mention that one can arrive at an effective 2D theory, similar to previous $\text{SU}(2)\times \text{SU}(2)$ approaches to spin-valley physics in Si \cite{Yannouleas2022}, by projecting the Hamiltonian in Eq. (\ref{Hp}) onto the subspace containing the ground ($\ell = 0$) and excited ($\ell = 1$) valleys.


\subsection{Summary of calculation procedure}
To conclude this section, we present a brief summary of the procedure used in a typical calculation, like those reported in Sec.~\ref{Results}. First, we specify a Ge concentration profile as a function of layer index $n$. Second, we construct the subband Hamiltonian $H_0^{(z)}$ using Eq.~(\ref{Ho}). 
Third, we diagonalize the subband Hamiltonian to obtain a set of eigenstates $\left\{\ket{\varphi_\ell}\right\}$. Finally, we construct the $H_R^{(z)}$ and $H_D^{(z)}$ matrices in Eqs.~(\ref{HR}) and (\ref{HD}) and calculate the matrix elements in Eqs.~(\ref{alphaElems}) and (\ref{betaElems}), which yields the spin-orbit coupling coefficients.

\section{Numerical results} \label{Results}
In this section, we present our numerical results. We first present in Sec.~\ref{ConventionalResults} spin-orbit coupling results for a ``conventional'' Si/SiGe quantum well system \textit{without} Ge oscillations included in the quantum well region. These results serve as a baseline for comparison. Next, we provide spin-orbit coupling results in Sec.~\ref{WiggleWellResults} for the Si/SiGe quantum well system \textit{with} Ge oscillations included in the quantum well region, namely, a wiggle well. In this case, we observe significant enhancement of the Dresselhaus spin-orbit coupling for appropriate Ge concentration oscillation wavelengths $\lambda$. In Sec.~\ref{RabiResults}, we show that the enhanced spin-orbit coupling allows for fast Rabi oscillations using EDSR. Finally, we study in Sec.~\ref{DisorderResults} the impact of alloy disorder on the spin-orbit coupling. 

\subsection{Spin-orbit coupling in Si/SiGe quantum wells \textit{without} Ge concentration oscillations} \label{ConventionalResults}

We first calculate spin-orbit coupling for the conventional Si/SiGe quantum well shown in Fig.~\ref{FIG1}(c). We assume barrier regions with a uniform Ge concentration of $n_\text{Ge,bar} = 30 \%$ and a quantum well width of $L_z \approx 20~\text{nm}$, consisting of an even number of atomic layers; the latter value was chosen to ensure that the wave functions have negligible weight at the bottom barrier except in the limit of very weak electric fields. In addition, the interfaces are given a nonzero width of $L_\text{int}\approx 0.95~\text{nm}$ ($7$ atomic layers), in which the Ge concentration linearly interpolates between values appropriate for the barrier and well regions. Such finite-width interfaces occur in realistic devices, and are known to significantly impact important properties of the quantum well such as the valley splitting~\cite{Wuetz2021}.

\begin{figure}[t]
\begin{center}
\includegraphics[width=0.48\textwidth]{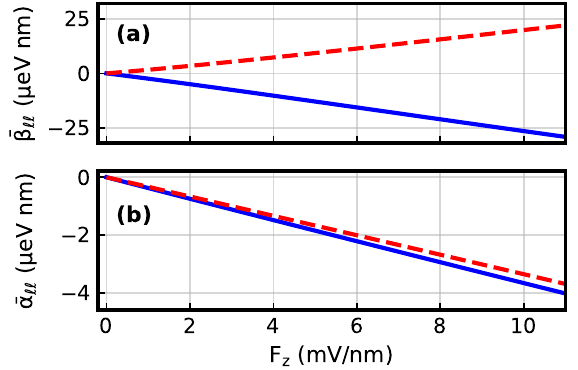}
\end{center}
\vspace{-0.5cm}
\caption{Diagonal Dresselhaus $\bar{\beta}_{\ell \ell}$ (a) and Rashba $\bar{\alpha}_{\ell \ell}$ (b) spin-orbit coupling coefficients as a function of vertical electric field $F_z$, for the ground (solid blue, $\bar{\beta}_{00}$ and $\bar{\alpha}_{00}$) and excited (dashed red, $\bar{\beta}_{11}$ and $\bar{\alpha}_{11}$) valleys of the ``conventional'' Si/SiGe quantum well shown in Fig.~\ref{FIG1}(c). Notice that the Dresselhaus coefficients $\bar{\beta}_{\ell \ell}$ are $\sim\!7$ times larger in magnitude than the Rashba coefficients $\bar{\alpha}_{\ell \ell}$.}
\label{FIG3}
\vspace{-1mm}
\end{figure}

The diagonal Dresselhaus and Rashba spin-orbit coupling coefficients are calculated for the ground (solid blue, $\bar{\beta}_{00}$ and $\bar{\alpha}_{00}$) and excited (dashed red, $\bar{\beta}_{11}$ and $\bar{\alpha}_{11}$) valley states, and are plotted in Figs.~\ref{FIG3}(a) and \ref{FIG3}(b) as a function of vertical electric field $F_z$. 
At zero field, $F_z = 0$, the diagonal spin-orbit coupling coefficients in Fig.~\ref{FIG3} all vanish, as consistent with the system being inversion symmetric \cite{Nestoklon2008,Prada2011}. 
By turning on the electric field we break the structural inversion symmetry, and the resulting spin-orbit coefficients vary linearly over the entire field range considered here. 
Note again that Dresselhaus spin-orbit coupling in SiGe requires the presence of a broken structural inversion symmetry~\cite{Ivchenko1996,Nestoklon2008,Prada2011,Vervoort1997}, in contrast with GaAs, which requires a broken bulk inversion symmetry~\cite{Dresselhaus1955}.  As consistent with previous studies~\cite{Nestoklon2006,Nestoklon2008,Veldhorst2015,Ruskov2018}, the Dresselhaus coefficients of the ground and excited valleys are found to be approximately opposite in sign, with the Dresselhaus coefficient of the excited valley being slightly smaller in magnitude. Additionally, the diagonal Rashba matrix elements are seen to be much smaller in magnitude than the Dresselhaus elements. Here, we find that $|\bar{\beta}_{\ell \ell}| \approx 7 |\bar{\alpha}_{\ell \ell}|$ for both low-energy valleys ($\ell = 0,1$). 
In addition, the magnitudes of the spin-orbit elements are found to quantitatively agree with Ref.~\cite{Nestoklon2008}, in the large electric field regime. (The system studied in Ref.~\cite{Nestoklon2008} assumed a narrower quantum well, resulting in different behavior at low electric fields.) We note, however, that the Rashba coefficients of the ground and excited valley states were found to have opposite signs in Ref.~\cite{Nestoklon2008}, while we find them to have the same sign here. This difference in our results occurs because we have used a softened interface where the Ge concentration interpolates between the barrier and well regions over a finite width, whereas Ref.~\cite{Nestoklon2008} used a completely sharp interface. We have numerically confirmed our results for the diagonal spin-orbit matrix elements using the computational scheme of Ref.~\cite{Nestoklon2008}, which is unrelated to our scheme, summarized in Eqs.~(\ref{alphaElems}) and (\ref{betaElems}).

\subsection{Spin-orbit coupling in Si/SiGe quantum wells \textit{with} Ge concentration oscillations} \label{WiggleWellResults}

We now calculate spin-orbit coefficients for the wiggle well geometry shown in Fig.~\ref{FIG1}(b), with a sinusoidally varying Ge concentration ranging from $n_{\text{Ge}} = 0\%$ to a maximum amplitude of $n_{\text{Ge}} = 10\%$, and an oscillation wavelength of $\lambda$. These parameters were chosen to match those of an experimental device reported in Ref.~\cite{McJunkin2021}. Here, the system parameters, $n_\text{Ge,bar}$, $L_z$, and $L_\text{int}$ are the same as before, and we apply an electric field of $F_z = 10~\text{mV/nm}$.

\begin{figure}[t]
\begin{center}
\includegraphics[width=0.48\textwidth]{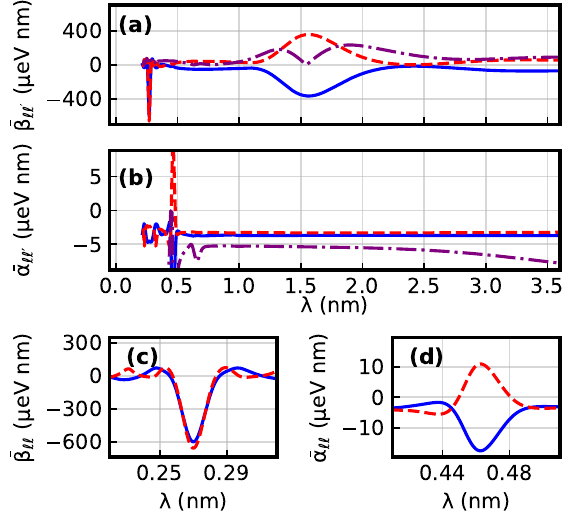}
\end{center}
\vspace{-0.5cm}
\caption{Dresselhaus $\bar{\beta}_{\ell\ell^\prime}$ (a) and Rashba $\bar{\alpha}_{\ell\ell^\prime}$ (b) spin-orbit matrix elements as a function of Ge oscillation wavelength $\lambda$ for an electric field of strength $F_z = 10~\text{mV/nm}$. Coefficients for the ground and excited valleys are shown in blue (solid, $\bar{\beta}_{00}$ and $\bar{\alpha}_{00}$) and red (dashed, $\bar{\beta}_{11}$ and $\bar{\alpha}_{11}$), respectively.  The off-diagonal elements ($\bar{\beta}_{01}$ and $\bar{\alpha}_{01}$), often referred to as the spin-valley coefficients, are shown as purple dashed-dotted lines. A wide bump centered at $\lambda \approx 1.57~\text{nm}$ occurs in (a), corresponding to a dramatically enhanced Dresselhaus spin-orbit coupling. At the center of the bump, $|\bar{\beta}_{\ell\ell}|$ is $\sim\!15$ times larger than the ``conventional'' Si/SiGe system at the same electric field [See Fig.~\ref{FIG3}(a).] Narrow bumps for the diagonal Dresselhaus and Rashba coefficients at small $\lambda$ values are shown in (c) and (d). The corresponding Ge concentration profile is shown in Fig.~\ref{FIG1}(b), where the average Ge concentration in the quantum well region is $\bar{n}_{\text{Ge}} = 5\%$.}
\label{FIG4}
\vspace{-1mm}
\end{figure}

\begin{figure}[t]
\begin{center}
\includegraphics[width=0.48\textwidth]{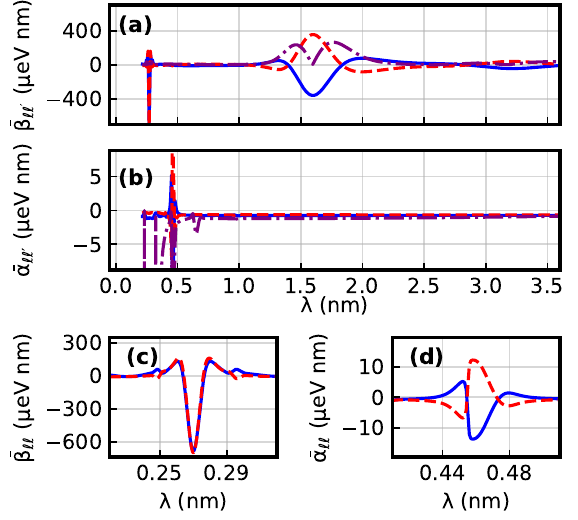}
\end{center}
\vspace{-0.5cm}
\caption{Same as Fig.~\ref{FIG4} except with an electric field strength of $F_z = 2~\text{mV/nm}$. Notice that the magnitude of $\bar{\beta}_{\ell\ell}$ at the peaks of the bumps are nearly the same as in Fig.~\ref{FIG4}, but that the features are narrower than in Fig.~\ref{FIG4}.}
\label{FIG5}
\vspace{-1mm}
\end{figure}

The resulting diagonal Dresselhaus $\bar{\beta}_{\ell \ell}$ and Rashba $\bar{\alpha}_{\ell \ell}$ spin-orbit coefficients for the ground (solid blue, $\bar{\beta}_{00}$ and $\bar{\alpha}_{00}$) and excited (dashed red, $\bar{\beta}_{11}$ and $\bar{\alpha}_{11}$) valley states are shown in Figs.~\ref{FIG4}(a) and \ref{FIG4}(b) as a function of the Ge oscillation wavelength $\lambda$. We find that the spin-orbit coefficients vary nontrivially with the choice of oscillation wavelength $\lambda$, exhibiting bumps where the coefficients are significantly enhanced. In particular, we observe a broad bump in Fig.~\ref{FIG4}(a), centered at $\lambda \approx 1.57~\text{nm}$, in which the Dresselhaus spin-orbit coefficient reaches the values $|\bar{\beta}_{00}| \approx 365~\mu\text{eV} \cdot \text{nm}$ for the ground valley and $|\bar{\beta}_{11}| \approx 357~\mu\text{eV} \cdot \text{nm}$ for the excited valley. 
Note that this peak location corresponds to the same wavelength as the long-period wiggle well predicted in Refs.~\cite{McJunkin2021,Feng2022} to enhance the valley splitting. (Note that a slightly different $\lambda$ value was predicted in Refs. \cite{McJunkin2021,Feng2022}, due to the valley minima residing at slightly different momenta $\pm k_0$ compared to our model.) 
Similar to the conventional Si/SiGe quantum well shown in Fig.~\ref{FIG4}, the ground and excited valleys have approximately opposite Dresselhaus coefficients for the range of $\lambda$ values considered here, except at very small $\lambda$. At the peak of the bump, however, the diagonal Dresselhaus coefficients $\bar{\beta}_{\ell \ell}$ for the wiggle well are $\sim\!15$ times larger than those of the conventional system at the same electric field strength. [See Fig.~\ref{FIG3}(a) at $F_z = 10~\text{mV/nm}$.] Moreover, the region of enhancement appears very broad, with a full-width at half-maximum of $\Delta \lambda = 0.55~\text{nm}$. In contrast to the diagonal Dresselhaus coefficients $\bar{\beta}_{\ell\ell}$, the diagonal Rashba coefficients $\bar{\alpha}_{\ell\ell}$ show no enhancement features except at small wavelengths. Indeed, for $\lambda \gtrsim 0.5~\text{nm}$, the Rashba coefficients are essentially independent of $\lambda$ with $\bar{\alpha}_{00} \approx -3.7~\mu\text{eV} \cdot \text{nm}$ and $\bar{\alpha}_{11} \approx -3.3~\mu\text{eV} \cdot \text{nm}$ for the ground and excited valleys, respectively. Comparing this to the results of the conventional system in Fig.~\ref{FIG3}(b) for the same electric field, we see that the Rashba elements are nearly identical in the two cases. Evidently, the Rashba coefficients are unaffected by the Ge concentrations oscillations for all but the shortest $\lambda$ values. 

We also plot the off-diagonal spin-orbit coupling elements $(\bar{\beta}_{01}$ and $\bar{\alpha}_{01})$ as purple dashed-dotted lines in Figs.~\ref{FIG4}(a) and ~\ref{FIG4}(b). Note that these quantities are often referred to as spin-valley coupling. Interestingly, the off-diagonal Dresselhaus coefficient $\bar{\beta}_{01}$ vanishes at the center of the $\lambda \approx 1.57~\text{nm}$ feature. Except at this particular wavelength for the Dresselhaus spin-orbit coupling, the off-diagonal spin-orbit coefficients are comparable in size to the diagonal coefficients.

To illustrate the features at small wavelengths, we plot the diagonal Dresselhaus and Rashba coefficients over selected, narrow ranges of $\lambda$ in Figs.~\ref{FIG4}(c) and \ref{FIG4}(d). Here we observe an enhancement in the Dresselhaus coefficient $\bar{\beta}_{\ell\ell}$ centered at $\lambda = a/2 \approx 0.27~\text{nm}$, where $a$ is the size of the cubic unit cell in Fig.~\ref{FIG2}(a) with an amplitude about twice that of the $\lambda = 1.57~\text{nm}$ peak. This $\lambda$ value corresponds to having a Ge concentration profile that alternates on every other atomic layer, essentially transforming the diamond lattice of Si into a zinc-blende lattice. In this case, the system can be thought of as a III-V semiconductor with the cation and anion corresponding to different concentrations of Ge. Interestingly, we find that the ground and excited valleys have the same sign of $\bar{\beta}_{\ell\ell}$ here, in contrast to the bump centered at $\lambda \approx 1.57~\text{nm}$. We also observe a enhancement in the Rashba coefficient $\bar{\alpha}_{\ell\ell}$ near $\lambda = 0.46~\text{nm}$ as shown in Fig.~\ref{FIG4}(d). However, note that the magnitude of this peak is still more than an order of magnitude smaller than the $\bar{\beta}_{\ell\ell}$ peaks. Indeed, the magnitude of this peak is even smaller than the Dresselhaus coefficients $\bar{\beta}_{\ell\ell}$ shown in Fig.~\ref{FIG3}(a) for the conventional Si/SiGe system at the same electric field strength. 
Finally, we note that these small-$\lambda$ bumps have a much narrower width than the $\bar{\beta}_{\ell\ell}$ bump at $\lambda = 1.57~\text{nm}$, which has important consequences for practical applications.

To study the tunability of the spin-orbit physics, we also calculate the spin-orbit coefficients of a wiggle well for a weaker vertical electric field. These results are shown in Fig.~\ref{FIG5}, which is the same system as Fig.~\ref{FIG4}, except with $F_z = 2~\text{mV/nm}$. The results are qualitatively similar to the stronger electric field case. However, there exists an important quantitative similarity and difference between the two cases, which we address in the following two paragraphs.

The remarkable similarity is that the Dresselhaus coefficients $\bar{\beta}_{\ell\ell}$ at the peaks of the bumps are nearly identical in the two cases. Note that this is true for both the $\lambda = 1.57~\text{nm}$ and $\lambda = 0.27~\text{nm}$ bumps. Evidently, at the center of the bumps, the electric field plays a minor role. This is in stark contrast with the Dresselhaus coefficients for the conventional system, where the diagonal Dresselhaus coefficients are proportional to $F_z$ as shown in Fig.~\ref{FIG3}(a). This highlights the important difference first discussed in the Introduction between spin-orbit coupling in conventional Si/SiGe quantum wells and wiggle wells, namely, that the conventional system requires the presence of an interface and structural asymmetry, while the wiggle well fundamentally does not. In the latter case, the spin-orbit coupling is an intrinsic property of the \textit{bulk} system. Indeed, we have checked that the Dresselhaus spin-orbit coupling persists in a wiggle well of wavelength $\lambda = 1.57~\text{nm}$, in the absence of an interface, by calculating the spin-orbit coefficients for a system without barriers, but instead immersed in a harmonic potential, $V_n = V_0 z_n^2$, obtaining similar results.

The main difference between the two cases is that the features are narrower in $\lambda$ space for the weak electric field of Fig.~\ref{FIG5}, compared to the strong electric field of Fig.~\ref{FIG4}. This represents an important advantage of strong electric fields for the wiggle well system, since it can be challenging to grow heterostructures with perfect oscillation periods. We conclude that stronger electric fields provide more reliable access to the enhanced spin-orbit coupling provided by Ge concentration oscillations, since the control of $\lambda$ does not need to be as precise during the growth process. The reason for narrower features in weaker electric field will be explained in Sec.~\ref{ExtPWB}. Finally, we note that stronger electric fields also provide larger valley splittings~\cite{Boykin2004a}.

\subsection{Electric dipole spin resonance} \label{RabiResults}

\begin{figure}[t]
\begin{center}
\includegraphics[width=0.48\textwidth]{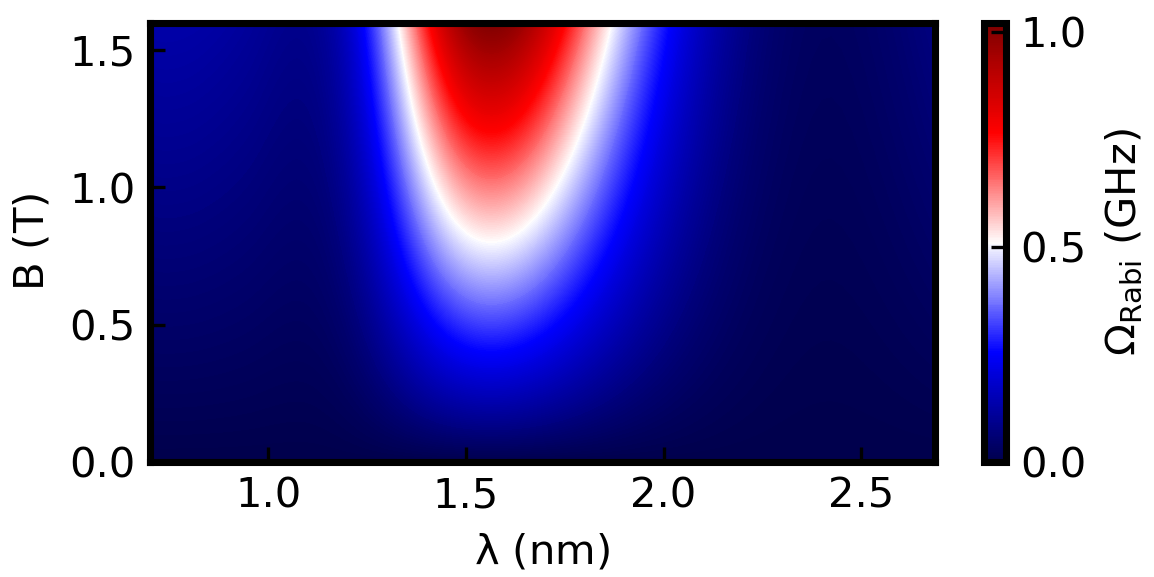}
\end{center}
\vspace{-0.5cm}
\caption{Electric dipole spin resonance (EDSR) Rabi frequency $\Omega_\text{Rabi}$ as a function of magnetic field $B$ and Ge concentration oscillation wavelength $\lambda$ for a quantum dot with confinement energy $\hbar \omega = 1~\text{meV}$ and an in-plane AC electric field amplitude $F_x = 10^{-2}~\text{mV/nm}$. The vertical electric field is $F_z = 10~\text{mV/nm}$.} 
\label{FIG6}
\vspace{-1mm}
\end{figure}

Aside from being of interest from a purely scientific standpoint, the presence of spin-orbit coupling can be exploited to perform gate operations within a quantum computation context. In particular, electric dipole spin resonance (EDSR)  is a powerful technique to manipulate individual spins through all-electrical means~\cite{Golovach2006,Rashba2008}. Here, we calculate the EDSR Rabi frequency for a single electron in a quantum dot embedded in a wiggle well. As shown in Ref.~\cite{Rashba2008}, applying an AC, in-plane electric field of amplitude $F_x$ with frequency $\omega_d$ across a quantum dot with spin-orbit coupling leads to an effective AC in-plane magnetic field. For valley $\ell$, the magnitude of this effective AC magnetic field is given by 
\begin{equation}
    B_{\text{eff}}(t)  = \frac{2eF_x}{\hbar \omega} \left(\frac{\omega_d}{\omega}\right)
    \frac{\sqrt{\bar{\beta}^2_{\ell \ell} + \bar{\alpha}^2_{\ell \ell}}}{\textsl{g} \mu_B} \sin\left(\omega_d t\right),
\end{equation}
where $\hbar \omega$ is the level spacing characteristic of the dot's harmonic confinement potential, $\textsl{g}$ is the $\textsl{g}$-factor, and $\mu_B$ is the Bohr magneton. Note that for $|\bar{\beta}_{\ell \ell}| \gg |\bar{\alpha}_{\ell \ell}|$, as is the case for Si/SiGe systems, the effective AC magnetic field is parallel to the in-plane electric field $F_x$. Applying a static, out-of-plane magnetic field $B$, the effective Hamiltonian for the quantum dot restricted to the orbital ground state is then
    $H_{\text{eff}}(t) = \frac{1}{2} \textsl{g} \mu_B 
    \left[
    B \sigma_z + B_{\text{eff}}(t) \sigma_x
    \right]$.
For a system initialized in the spin-$\uparrow$ ground valley state and driven at the resonance frequency $\omega_d = \textsl{g} \mu_B B / \hbar$ set by the external magnetic field, the probability of finding the electron in the spin-$\downarrow$ state is given by
$P_\downarrow(t) = \sin^2(\Omega_\text{Rabi} t/2)$~\cite{Townsend2012}, where the Rabi frequency is found to be
\begin{equation}
    \Omega_\text{Rabi} = 
    \frac{eF_x \textsl{g} \mu_B\sqrt{\bar{\beta}^2_{\ell \ell} + \bar{\alpha}^2_{\ell \ell}}}{\hbar(\hbar \omega)^2} 
    B.
\end{equation}
The EDSR Rabi frequency $\Omega_\text{Rabi}$ of a quantum dot in a wiggle well is plotted in Fig.~\ref{FIG6} as a function of magnetic field and Ge oscillation wavelength $\lambda$ for the realistic parameters of $\hbar \omega = 1~\text{meV}$ and $F_x = 10^{-2}~\text{mV/nm}$. We see that for oscillation wavelengths near the peak of the the spin-orbit enhancement at $\lambda \approx 1.57~\text{nm}$, we can obtain Rabi frequencies of $100$'s of MHz for moderate magnetic field strengths. Indeed, a Rabi frequency of $\Omega_\text{Rabi} = 1~\text{GHz}$ is achieved at the peak of the spin-orbit enhancement for a magnetic field $B \approx 1.5~\text{T}$. We therefore conclude that including Ge concentration oscillations in Si/SiGe quantum wells enables a dramatic speedup of single qubit gates using EDSR.

\subsection{Impact of alloy disorder} \label{DisorderResults}
 
SiGe is a random alloy, and the resulting alloy disorder is known to have an effect on quantum well properties such as valley splitting, both for conventional quantum wells~\cite{Wuetz2021} and wiggle wells~\cite{McJunkin2021,Feng2022}. It is therefore important to explore the effects of alloy disorder on spin-orbit coupling.
Recall from Sec.~\ref{VCModel} that we have employed a virtual crystal approximation in our model, which averages over all possible alloy realizations. While providing tractability to our calculations, this approximation ignores the fluctuations arising from the random nature of the Ge atom arrangements in the SiGe alloy. 
In this section, we explain how Ge concentration fluctuations can be reintroduced into our model, to explore the effects of alloy disorder.

A full 3D calculation including alloy disorder is computationally expensive due to the loss of translation invariance, and is beyond the scope of this work. We can still, however, include the effects of alloy disorder approximately within our 1D effective model by allowing for fluctuations in the Ge concentration in each atomic layer. This is accomplished using the procedure described in Ref.~\cite{Wuetz2021}, which can be summarized as follows. First, we assume a dot of radius $a_\text{dot} = \sqrt{\hbar/m_\parallel \omega} = 20~\text{nm}$ in the plane of the quantum well, where $m_\parallel = 0.19m_e$ is the in-plane effective mass and $\hbar\omega = 1~\text{meV}$ is the orbital excitation energy characterizing the parabolic confinement of the dot. We then calculate the \textit{effective} Ge concentration $n_{\text{Ge},n}^{\text{eff}}$ in layer $n$ of our disordered system by counting the number of Ge atoms \textit{within} our dot. Here, the probability of any given atom in layer $n$ being a Ge atom is $n_{\text{Ge},n}$, where $n_{\text{Ge},n}$ is the average germanium concentration throughout the entire layer $n$, and the number of atoms in our dot is $N_{\text{eff}} = 4\pi a_\text{dot}^2/a^2 \approx 17100$, where $a=0.543~\text{nm}$ is the cubic lattice constant of Si. The effective Ge concentration $n_{\text{Ge},n}^{\text{eff}}$ in layer $n$ can then be drawn from the distribution $n_{\text{Ge},n}^{\text{eff}} = N_{\text{eff}}^{-1} \text{Binom}(N_{\text{eff}},n_{\text{Ge},n})$, where $\text{Binom}(n,p)$ is the binomial distribution with $n$ trials and probability of success $p$. In the limit of $N_{\text{eff}} \rightarrow \infty$, the resulting, randomized effective Ge concentration $n_{\text{Ge},n}^{\text{eff}}$ approaches the ideal Ge concentration $n_{\text{Ge},n}$, but for smaller dots, fluctuations from this ideal limit become more pronounced. We then calculate the spin-orbit coefficients for our effective 1D model as in previous sections but with the Ge concentration profile given by $n_{\text{Ge},n}^{\text{eff}}$ instead of $n_{\text{Ge},n}$. Note that this method of including alloy disorder in the 1D effective model was shown in Ref.~\cite{Wuetz2021} to yield valley splitting distributions in good agreement with 3D calculations.

\begin{figure}[t]
\begin{center}
\includegraphics[width=0.48\textwidth]{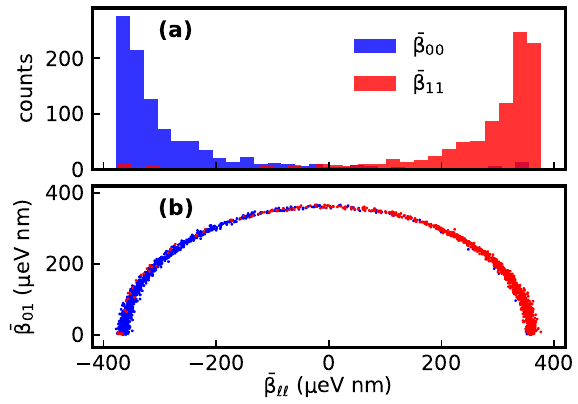}
\end{center}
\vspace{-0.5cm}
\caption{(a) Distribution of the diagonal Dresselhaus coefficients $\bar{\beta}_{\ell\ell}$ of the wiggle well when including alloy disorder as described in the main text. System parameters are $\lambda = 1.57~\text{nm}$ and $F_z = 10~\text{mV/nm}$. The data includes $1000$ alloy realizations. We see that the spin-orbit coupling enhancement is robust against alloy fluctuations with $86\%$ of alloy realizations having $|\bar{\beta}_{00}| > 200~\mu\text{eV}\cdot\text{nm}$. (b) Scatter plot of the diagonal [$\bar{\beta}_{00}$ (blue) and $\bar{\beta}_{11}$ (red)] and off-diagonal ($\bar{\beta}_{01}$) spin-orbit coefficients of all alloy realizations. This result indicates that the main effect of the alloy disorder is to mix the ground and excited valley states. 
}
\label{FIG7}
\vspace{-1mm}
\end{figure}

We now calculate the diagonal Dresselhaus coefficients $\bar{\beta}_{\ell\ell}$ for the same Si/SiGe system as Fig.~\ref{FIG4}, with Ge concentration oscillations of wavelength $\lambda = 1.57~\text{nm}$, but now with Ge concentration fluctuations included. Note that this $\lambda$ value corresponds to the peak of the main spin-orbit enhancement bump in Fig.~\ref{FIG4}(a). The distribution of the spin-orbit coefficients is plotted in Fig.~\ref{FIG7}(a) for $1000$ random-alloy realizations, for the ground (blue, $\bar{\beta}_{00}$) and excited (red, $\bar{\beta}_{11}$) valleys. Unsurprisingly, we see that the alloy fluctuations affect the spin-orbit coefficients, with the ground valley coefficient spanning the range $-377 < \bar{\beta}_{00} <355~\mu\text{eV}\cdot\text{nm}$. The distributions are highly peaked, however, near the \textit{bare} values $\bar{\beta}_{00} = -365~\mu\text{eV}\cdot\text{nm}$ and $\bar{\beta}_{11} = 357~\mu\text{eV}\cdot\text{nm}$ of the disorder-free system (see Fig.~\ref{FIG4}). Indeed, we find that $86\%$ of the alloy realizations have $|\bar{\beta}_{00}| > 200~\mu\text{eV}\cdot\text{nm}$. We therefore conclude that the spin-orbit enhancement arising from the Ge concentration oscillations within wiggle wells is robust against alloy fluctuations. 

We gain further insight into the effects of the alloy fluctuations by studying the \textit{inter}-valley spin-orbit coefficient $\bar{\beta}_{01}$. Figure~\ref{FIG7}(b) presents a scatter plot showing both the diagonal ($\bar{\beta}_{00}$ and $\bar{\beta}_{11}$) and off-diagonal ($\bar{\beta}_{01}$) coefficients of all $1000$ alloy realizations. Interestingly, all realizations yield coefficients that land in a narrow semi-circular region of parameter space. This is an indication that the alloy disorder is essentially mixing the original ground and excited valley states of the system without disorder. Note that perturbations arising from higher orbital subbands are insignificant, except for inducing a small width to the distribution. This is not unexpected since the energy difference between the ground and excited valley states is $<\!1~\text{meV}$, while the lowest orbital excitation energy is $\gtrsim \!20~\text{meV}$. 

\begin{figure}[t]
\begin{center}
\includegraphics[width=0.48\textwidth]{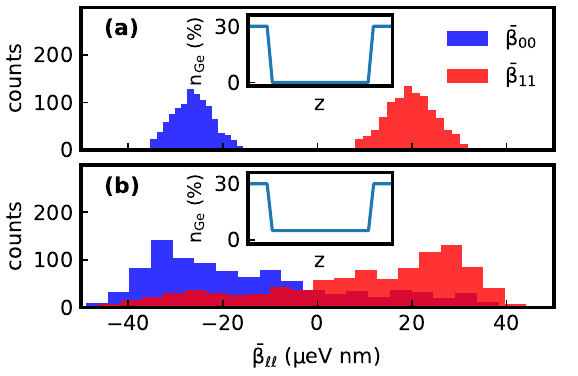}
\end{center}
\vspace{-0.5cm}
\caption{(a) Distribution of the diagonal Dresselhaus coefficients $\bar{\beta}_{\ell\ell}$ of a conventional Si/SiGe system with a pure Si quantum well when including alloy disorder as described in the main text. The electric field is $F_z = 10~\text{mV/nm}$. (b) Same as (a) except that $n_{\text{Ge},n} = 5\%$ throughout the well region. Insets show the $n_{\text{Ge},n}$ profiles. We see that inclusion of Ge inside of the well region widens the distribution of $\bar{\beta}_{\ell\ell}$ but does not on average increase its magnitude.}
\label{FIG8}
\vspace{-1mm}
\end{figure}

For completeness, we also calculate the distribution of Dresselhaus spin-orbit coefficients in conventional Si/SiGe quantum wells that include alloy fluctuations. These results are shown in Fig.~\ref{FIG8}(a) for a system with no Ge in the well region and electric field $F_z = 10~\text{mV/nm}$. Here, the Ge concentration profile is shown in the inset. Again, we obtain $\bar{\beta}_{\ell\ell}$ distributions centered at the same values as the disorder-free system. (See Fig.~\ref{FIG3}, with $F_z=10$~mV/nm.) Here, however, the spread is significantly narrower than the wiggle well results shown in Fig.~\ref{FIG7}(a), with the full-width-at-half-maximum being $\sim 10~\mu\text{eV}\cdot\text{nm}$. This is because no alloy fluctuations occur in the well region where the majority of the wave function resides. We also consider a system with $n_\text{Ge} = 5\%$ distributed uniformly throughout the well region, taking into account the effects of random alloy fluctuations. [Here, we do not include intentional Ge concentration oscillations; the resulting Ge concentration profile is shown in the inset of Fig.~\ref{FIG8}(b).] Note that this system contains the same amount of Ge in the well region as the wiggle well system shown in Fig.~\ref{FIG7}. Unsurprisingly, we find that the distribution of the $\bar{\beta}_{\ell\ell}$ coefficients spreads considerably, compared to the results in Fig.~\ref{FIG8}(a), due to the presence of alloy fluctuations inside of the well. Importantly, however, a uniform Ge concentration in the well region does not on average increase the spin-orbit coupling, in contrast to the effect on valley splitting. Indeed, we find that $\left<|\bar{\beta}_{00}|\right> = 26~\mu\text{eV}\cdot\text{nm}$ for the data in Fig.~\ref{FIG8}(a), while $\left<|\bar{\beta}_{00}|\right> = 22~\mu\text{eV}\cdot\text{nm}$ for the data in Fig.~\ref{FIG8}(b). This highlights that fact that in order to enhance spin-orbit coupling, it is not enough to simply include Ge in the well region, but rather the Ge concentration must oscillate with the appropriate wavelength $\lambda$. We note that this result is consistent with the experimental observation that the $\textsl{g}$-factor measured in uniform Si$_{1-x}$Ge$_x$ alloys is only slightly altered by changing the Ge concentration \cite{Jantsch2002, Malissa2004}.

\section{Mechanism behind the spin-orbit enhancement} \label{Mechanism}

Having shown through simulations that the inclusion of Ge concentration oscillations of an appropriate wavelength can significantly enhance the spin-orbit coupling, the natural question is what mechanism leads to this enhancement? In this section, we provide an explanation of this mechanism. To begin, we describe in Sec.~\ref{SimpModel} a simplified version of our model that allows for easier understanding of the spin-orbit enhancement mechanism. Next, we study in Sec.~\ref{RealSpaceSec} the real-space representation of the ground valley wave function in the absence and presence of Ge concentration oscillations. We find that a real-space picture is inadequate in explaining the spin-orbit enhancement. We therefore study in Sec.~\ref{ExtPWB} 
the structure of the wave function in momentum space. We show how the combination of the oscillating potential produced by the Ge concentration oscillations and the selection rules of Dresselhaus spin-orbit coupling leads to the spin-orbit coupling enhancement.  

\subsection{Simplified Model} \label{SimpModel}

To focus on the essential physics for the spin-orbit coupling enhancement, we use in this section a model for SiGe alloys that is slightly simplified compared to the model presented in Sec.~\ref{Model} and used in our numerical calculations in Sec.~\ref{Results}. 
In this model, Ge atoms are assumed to be identical to Si atoms, except for their orbitals being shifted up in energy by a constant, i.e. $\bar{\varepsilon}_{\bar{\nu}}^{(\text{Ge})} = \bar{\varepsilon}_{\bar{\nu}}^{(\text{Si})} + E_\text{Ge}$ where $E_\text{Ge} = 0.8~\text{eV}$ is the extra energy of every Ge orbital. This is meant to capture at the simplest level that inclusion of Ge increases the energy of the conduction band minima. In particular, the chosen value produces a band offset of $0.24~\text{eV}$ between a pure Si region and a barrier region with a nominal 30\% Ge concentration. 
Note, however, that the precise value is not important since we are only using this simplified model to understand the spin-orbit enhancement mechanism, leaving quantitative questions to the more accurate model of Sec.~\ref{Model}.  In addition, we also neglect the effects of strain, such that $\bar{C}^{(n)} \rightarrow 0$, since they are not crucial in understanding the spin-orbit enhancement mechanism. 
With these simplifications, the addition of Ge is equivalent to adding a term to the potential energy $V$ of a pure Si system. For simplicity, we define a new potential $\mathcal{V}_n = e F_z z_n + E_{\text{Ge}} n_{\text{Ge},n}$, that includes both the electric field $F_z$ and the energy shift from the Ge concentration $n_{\text{Ge},n}$ of the layer, and we let all orbitals energies take values appropriate for Si: $\bar{\varepsilon}_{\bar{\nu}}^{(n)} \rightarrow \bar{\varepsilon}_{\bar{\nu}}^{(\text{Si})}$. Here, $z_n = n a/4$ is the  $z$ coordinate of atomic layer $n$, and $a = 0.543~\text{nm}$ is the cubic lattice constant of Si. Importantly, in this simplified model, the onsite orbital energies and hopping matrices all lose their dependence on the layer index $n$,  $\left\{\bar{\varepsilon}_{\bar{\nu}}^{(n)}, \widetilde{T}^{(n)}_0, \widetilde{T}^{(n)}_R(\mathbf{k}_\parallel), \widetilde{T}^{(n)}_D(\mathbf{k}_\parallel), \Omega^{(n)}, \Phi^{(n)}\right\} \rightarrow \left\{\bar{\varepsilon}_{\bar{\nu}}^{(\text{Si})}, \widetilde{T}_0, \widetilde{T}_R(\mathbf{k}_\parallel), \widetilde{T}_D(\mathbf{k}_\parallel), \Omega, \Phi\right\}$. The Hamiltonian components then take the simplified forms
\begin{align}
    \begin{split}
    \mel{m \bar{\mu}}{H_0^{(z)}}{n \bar{\nu}} =& 
    ~\delta_{m\bar{\mu}}^{n\bar{\nu}}
    \left(\bar{\varepsilon}_{\bar{\nu}}^{(\text{Si})} + \mathcal{V}_n\right) 
    \\
    &+ \delta_{m}^{n+1} \Omega_{\bar{\mu} \bar{\nu}} 
    + \delta_{m}^{n-1} \Omega^{T}_{\bar{\mu} \bar{\nu}} \end{split}, \label{Ho2}\\
    \mel{m \bar{\mu}}{H_R^{(z)}}{n \bar{\nu}}=&
    \delta_{m}^{n+1} \Phi_{\bar{\mu} \bar{\nu}} 
    + \delta_{m}^{n-1} \Phi^{T}_{\bar{\mu} \bar{\nu}}, \label{HR2} \\
     \mel{m \bar{\mu}}{H_D^{(z)}}{n \bar{\nu}}=&
    (-1)^{n}  
    \left(\delta_{m}^{n+1} \Phi_{\bar{\mu} \bar{\nu}} 
    - \delta_{m}^{n-1} \Phi^{T}_{\bar{\mu} \bar{\nu}}\right). \label{HD2}
\end{align}
Note that the numerical values of the $\Omega$ and $\Phi$ matrices used in the above equations are specified in Appendix \ref{OmegaPhi}.

\subsection{Real-space wave functions} \label{RealSpaceSec}
\begin{figure}[t]
\begin{center}
\includegraphics[width=0.48\textwidth]{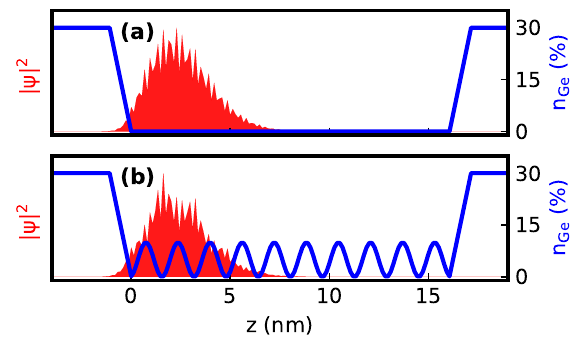}
\end{center}
\vspace{-0.5cm}
\caption{Comparison of the ground valley wave function for a ``conventional'' Si/SiGe system with a pure Si well region (a) and a wiggle well (b) with oscillation wavelength $\lambda = 1.62~\text{nm}$. Blue lines show the Ge concentration profiles while red filled in curves are the wave functions $|\psi|^2$. Note that we sum over orbital indices. The electric field is $F_z = 5~\text{mV/nm}$ in both (a) and (b). We see in (b) that the wave function is suppressed in regions of high Ge concentration, consistent with the fact that the conduction band minima are higher in energy in those regions.} 
\label{FIG9}
\vspace{-1mm}
\end{figure}

Let us now study the effects of the Ge concentration oscillations on the wave functions of the subband Hamiltonian $H_0^{(z)}$. To do so, we first calculate the ground valley wave function of a conventional Si/SiGe quantum well that has the Ge concentration profile shown as the blue line in Fig.~\ref{FIG9}(a). The ground valley wave function $|\psi|^2$ of the subband Hamiltonian $H_0^{(z)}$ is shown in red in Fig.~\ref{FIG9}(a), where the state is pushed up against the barrier-well interface by an electric field of strength $F_z = 5~\text{mV/nm}$. The wave function also exhibits fast oscillations characteristic of the superposition of the two valley minima~\cite{Friesen2007}. 
We next consider the same system, except with Ge concentration oscillations of wavelength $\lambda = 1.62~\text{nm}$ included in the well region, as shown by the blue line in Fig.~\ref{FIG9}(b). 
Comparing the two 
cases, we see that the wave function is suppressed in regions of high Ge concentration, as consistent with the fact that the conduction-band minima (and thus the local potential energies) are higher in energy in those regions. However, these qualitative observations do not directly explain the enhancement of Dresselhaus spin-orbit coupling observed in Figs.~\ref{FIG4} and \ref{FIG5}. 

\subsection{Momentum-space wave functions} \label{ExtPWB}

\begin{figure}[t]
\begin{center}
\includegraphics[width=0.48\textwidth]{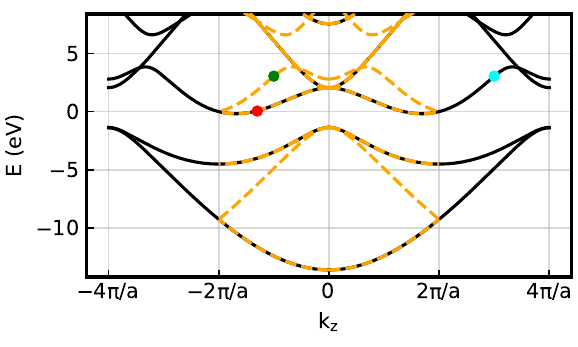}
\end{center}
\vspace{-0.5cm}
\caption{ The band structure of Si for $\boldsymbol{k}_\parallel = 0$. The band structure for the conventional Brillouin zone extending to $k_z = \pm 2 \pi/a$ is shown by dashed orange lines. The presence of band \textit{crossings} at the zone edge $k_z = \pm 2\pi/a$ indicates that the Brillouin zone can be enlarged. Indeed, the equivalence of the two sublattices of Si for $\mathbf{k}_\parallel = 0$ allows us to extend the Brillouin zone to $k_z = \pm 4\pi/a$. The band structure for the extended zone is shown as solid black lines. See the main text for details on the green, blue, and red points.}
\label{FIG10}
\vspace{-1mm}
\end{figure}

We gain a clearer understanding of the effects of the Ge concentration oscillations by studying the momentum-space representation of the ground valley. To begin, let us first investigate the band structure of Si at $\mathbf{k}_\parallel = 0$, which is shown as the orange dashed lines in Fig.~\ref{FIG10}. A central feature of the band structure is the presence of two degenerate valleys near zero energy that give rise to the ground and excited valley states in a quantum well. In addition, 
the valley minima are located 
only a short distance of $0.17(2\pi/a)$ away from the Brillouin zone edge at $k_z = \pm 2\pi/a$. Interestingly, we notice that at $k_z = \pm 2\pi/a$, there are no band anti-crossings (only crossings). 
This is an indication that the Brillouin zone can be enlarged for $\mathbf{k}_\parallel = 0$. 
Indeed, for the special case of $\mathbf{k}_\parallel = 0$, the two sublattice hopping matrices become equal, $\widetilde{T}_{+}(\mathbf{0}) = \widetilde{T}_{-}(\mathbf{0})$, and the primitive unit cell of the 1D chain shown in Fig. \ref{FIG2}(b) reduces from two sites to one. 
The Brillouin zone should therefore extend to $k_z = \pm 4\pi/a$ instead of $k_z = \pm 2\pi/a$. 
Here, we will refer to the Brillouin zone that extends to $k_z = \pm 4\pi/a$ as the \textit{extended zone}, while the zone extending only to $k_z = \pm 2\pi/a$ as the \textit{conventional zone}. 

To calculate the band structure in the extended zone, we define a plane wave basis state with momentum $k_z$ as
\begin{equation}
    \ket{k_z\bar{\nu}} = \frac{1}{\sqrt{N_z}}
    \sum_n e^{ik_z z_n} \ket{n \bar\nu}, \label{PWBS}
\end{equation}
where $N_z$ is the number of sites in the 1D chain and $-4\pi/a <k_z  \leq 4\pi/a$, where $k_z = 8\pi n/(N_z a)$. The subband Hamiltonian $H_0^{(z)}$ then has matrix elements,
\begin{equation}
    \begin{split}
    \mel{k_z \bar{\mu}}{H_0^{(z)}}{k_z^\prime \bar{\nu}} =& 
    \delta_{k_z}^{k_z^\prime} 
    \Big(
    \delta_{\bar{\mu}}^{\bar{\nu}}\bar{\varepsilon}_{\bar{\nu}}^{(\text{Si})} 
    +e^{-ik_z \frac{a}{4}}\Omega_{\bar{\mu}\bar{\nu}}
    +e^{ik_z \frac{a}{4}}\Omega^T_{\bar{\mu}\bar{\nu}}
    \Big) \\
    &+ \delta_{\bar{\mu}}^{\bar{\nu}}~ \widetilde{\mathcal{V}}(k_z-k_z^\prime)
    \end{split}, \label{HoMomentum}
\end{equation}
where $\widetilde{\mathcal{V}}(k_z-k_z^\prime)$ is the Fourier transform of the potential $\mathcal{V}$ and is given by $\widetilde{\mathcal{V}}(q_z) = N_z^{-1}\sum_{n} \exp(-i q_z z_n) \mathcal{V}_n$. Notice that in the absence of the potential [$\widetilde{\mathcal{V}}(q_z) \rightarrow 0$], $k_z$ is a good quantum number for $H_0^{(z)}$, and Eq.~(\ref{HoMomentum}) represents a Bloch Hamiltonian. The spectrum of $H_0^{(z)}$ for $\widetilde{\mathcal{V}}(q_z)= 0$ is shown as the solid black lines in Fig.~\ref{FIG10}, which we refer to as the extended band structure. Within the conventional zone, we see that the extended band structure aligns perfectly with half of the conventional (orange) bands, while the other half of the bands have been removed from the conventional zone and instead reside in the regions between $k_z = |2\pi/a|$ and $|4\pi/a|$. Furthermore, we observe that the conventional bands that do not match with the extended band structure can be made to match by shifting them by a reciprocal lattice vector $G_z = \pm 4\pi/a$ of the lattice containing two sites. For example, the green point at $k_z = -\pi/a$ in Fig.~\ref{FIG10} gets shifted to the blue point at $k_z = 3\pi/a$, which coincides with a band of the extended band structure. This is expected since the points $k_z$ and $k_z + G_z$ are equivalent from the point of view of the conventional zone \cite{Kittel2019}.

We stress that the extended band structure contains more information than the conventional band structure. Indeed, the extended scheme clarifies which states can be coupled by a given Fourier component $q_z$ of the potential $\widetilde{\mathcal{V}}(q_z)$. For example, let us consider what Fourier component of the potential could couple the states of the conventional band structure marked by the green and red points at $k_z = -\pi/a$ and $-1.3\pi/a$, respectively, in Fig.~\ref{FIG10}. Looking at the conventional band structure, one may initially believe the $\widetilde{\mathcal{V}}(-0.3 \pi/a)$ Fourier component 
could couple the states. Looking at these states in the extended band structure (red and blue points in Fig.~\ref{FIG10}), however, we immediately see that these states are instead coupled by the $\widetilde{\mathcal{V}}(3.7 \pi/a)$ Fourier component. 
This additional information will be crucial in understanding the enhanced spin-orbit coupling mechanism below.
\begin{figure}[t]
\begin{center}
\includegraphics[width=0.48\textwidth]{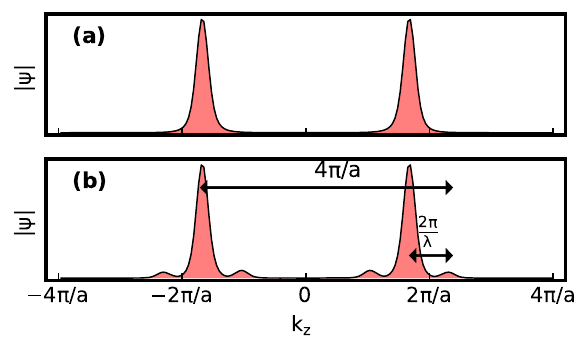}
\end{center}
\vspace{-0.5cm}
\caption{Wave-function profiles of the ground valley in the plane-wave representation. (a), (b) Correspond to Ge profiles (a) and (b) in Fig.~\ref{FIG9}, respectively. Both states show main peaks centered at the valley minima, $k_z \approx \pm 0.83(2\pi/a)$, as shown in Fig.~\ref{FIG10}. As shown in (b), however, the Ge concentration oscillations produce wave-function satellites a distance $2\pi/\lambda$ away from the main peaks. The wave vector from a main peak to the opposite outer satellite is $4\pi/a$ for this choice of $\lambda$.
Since Dresselhaus spin-orbit coupling connects locations in reciprocal space separated by $4\pi/a$, spin-orbit enhancement occurs when a satellite peak is separated from a main peak by $4\pi/a$.
}
\label{FIG11}
\vspace{-1mm}
\end{figure}

Returning to our comparison of the systems with and without Ge concentration oscillations, we now plot the ground valley wave functions in the plane-wave representation. These are shown in Figs.~\ref{FIG11}(a) and \ref{FIG11}(b) and correspond to the real-space wave functions shown in Figs.~\ref{FIG9}(a) and \ref{FIG9}(b), respectively. That is, we plot $(\sum_{\bar{\nu}} |\bra{k_z \bar{\nu}}\ket{\varphi_\ell}|^2)^{1/2}$ as a function of $k_z$ where $\ket{\varphi_\ell}$ is the ground valley wave function. 
As shown in Fig.~\ref{FIG11}(a), the ground valley wave function of the conventional Si/SiGe heterostructure consists of two peaks centered at $k_z \approx \pm 0.83 (2\pi/a)$. These coincide with the conduction-band minima in Fig.~\ref{FIG10}(a) as expected. Note that the positions of these minima in the band structure of Si depends on the exact tight binding parameters used, and differs slightly from other models in the literature. As shown in Fig.~\ref{FIG11}(b), the ground valley wave fuction in the presence of the Ge oscillations still has its main peaks 
but also includes surrounding satellite features. The location of these satellites with respect to the central peaks is determined by the wavelength of the Ge oscillations, with a peak-satellite separation of $2\pi/\lambda$, as shown in Fig.~\ref{FIG11}(b). 

The second step in explaining the enhanced spin-orbit coupling requires a mechanism for coupling different regions of the Brillouin zone.
We first express the matrix elements of the Rashba $H_R^{(z)}$ and Dresselhaus $H_D^{(z)}$ Hamiltonian components in the plane wave basis, giving
\begin{align}
    \mel{k_z \bar{\mu}}{H_R^{(z)}}{k_z^\prime \bar{\nu}} &= 
    \delta_{k_z}^{ k_z^\prime}
    \left(
    e^{-ik_z \frac{a}{4}} \Phi_{\bar{\mu}\bar{\nu}} +
    e^{ik_z \frac{a}{4}} \Phi^T_{\bar{\mu}\bar{\nu}}
    \right), \label{Cppw} \\
    \mel{k_z \bar{\mu}}{H_D^{(z)}}{k_z^\prime \bar{\nu}} &= 
    \delta_{|k_z - k_z^\prime|}^{4\pi/a}
    \Big(
    e^{-ik_z \frac{a}{4}} \Phi_{\bar{\mu}\bar{\nu}}
    -e^{ik_z \frac{a}{4}} \Phi^T_{\bar{\mu}\bar{\nu}}
    \Big). \label{Cmpw}
\end{align}
The key features to notice here are the selection rules between $k_z$ and $k_z^\prime$: while $H_R^{(z)}$ conserves $k_z$, 
$H_D^{(z)}$ couples states with momenta differing by $4\pi / a$. 
The latter result is obtained by Fourier transforming the $(-1)^{n}$ factor in Eq.~(\ref{HD2}), which itself is a manifestation of the alternating sign in front of $\widetilde{T}_D^{(n)}$ in Eq.~(\ref{Texpansion}). 
To our knowledge, these selection rules and their relation to Rashba and Dresselhaus spin-orbit couplings have not been noticed previously, although we speculate that they could be deduced from group-theory methods, such as the method of invariants \cite{Winkler2003,Li2011}. We emphasize, however, that the extended zone scheme is key to obtaining such results, since in the conventional zone scheme, $k_z$ values differing by $4\pi/a$ are equivalent, and cannot yield a selection rule like Eq.~(\ref{Cmpw}).
We also note that Eqs.~(\ref{Cppw}) and (\ref{Cmpw}) apply to any system with a diamond crystal structure.

The Dresselhaus momentum selection rule in Eq.~(\ref{Cmpw}) indicates that the spin-orbit coupling will be enhanced when wave-function peaks associated with two different valleys are separated by $4\pi/a$.
In Fig.~\ref{FIG11}(b), we see that this can only occur for coupling between a central valley peak and the outer satellite associated with the opposite valley.
Since the central valley peaks are located at $k_z \approx \pm 0.83 (2\pi/a)$, we see that the resonance condition for the oscillation wavelength to enhance the Dresselhaus spin-orbit is given by
$\lambda_\text{res} = 2.94 a = 1.59~\text{nm}$. Uncoincidentally, this wavelength value is very close to the peak of the bump centered at $\lambda \approx 1.57~\text{nm}$ of our numerical result in Fig.~\ref{FIG4}(a) where the Dresselhaus spin-orbit coupling is significantly enhanced. Note that similar considerations apply to the excited valley. 

Our understanding of the spin-orbit enhancement mechanism also explains the narrower features observed in Fig.~\ref{FIG5} for a weak vertical electric field $F_z = 2~\text{mV/nm}$, as compared to the wider features observed in Fig.~\ref{FIG4} for a strong electric field $F_z = 10~\text{mV/nm}$. Essentially, a weaker electric field produces wave functions with narrower features in the plane-wave representation than those from a strong electric field. 
These thinner features make it more difficult to satisfy the Dresselhaus resonance condition since the narrower peaks need to be situated more precisely in $k_z$ space.

Finally, we comment that the spin-orbit enhancement mechanism relies fundamentally on the degeneracy of the two $z$ valleys in the band structure of strained Si. Such a situation could not occur if, for example, Si was a direct bandgap semiconductor with a single non-degenerate valley at the $\Gamma$ point, since the key coupling in Fig.~\ref{FIG11}(b) occurs between the central peak of one valley and the outer satellite of the \textit{opposite} valley. Interestingly, in this case the valley degeneracy of Si can be considered as beneficial, while in other scenarios it is often considered to be problematic. In addition, the spin-orbit enhancement mechanism has similarities to holes in semiconductors, for which the enhancement arises due to degeneracy at the valence-band edge. In contrast to the valence-band case, however, which has degenerate bands at the same momentum, the degenerate $z$ valleys in Si require a periodic potential in the form of Ge concentration oscillations to enhance the coupling strength.

\section{Conclusions} \label{Conclusion}

We have shown that the inclusion of periodic Ge concentrations oscillations within the quantum well region of a Si/SiGe heterostructure leads to enhanced spin-orbit coupling when the oscillation wavelength $\lambda$ is properly chosen. Specifically, we find that the Dresselhaus spin-orbit coupling coefficient is enhanced by over an order of magnitude when $\lambda \approx 1.57~\text{nm}$, as shown in Figs.~\ref{FIG4} and \ref{FIG5}. We have provided a detailed explanation for this behavior: the Ge concentration oscillations produce wave function satellites in momentum space which can couple strongly to the valley minima through Dresselhaus spin-orbit coupling provided that the satellite-valley separation is approximately $4\pi/a$ in the extended Brillouin zone as shown in Fig.~\ref{FIG11}. 
Importantly, the region of enhancement in Fig.~\ref{FIG4} is quite wide in $\lambda$ space, which has the important implication that the wiggle well structure should allow for rather large growth errors in the Ge concentration profile while maintaining the enhanced spin-orbit effect. 
Additionally, our results indicate that the spin-orbit enhancement is robust against alloy disorder, as shown in Fig.~\ref{FIG7}. 

Enhancement of both the Dresselhaus and Rashba coefficients at smaller $\lambda$ values have also been found in Figs.~\ref{FIG4}(c) and \ref{FIG4}(d), although these bumps are much narrower in width than the $\lambda \approx 1.57~\text{nm}$ bump, making such structures more challenging to fabricate. Assuming that the wiggle well with $\lambda \approx 0.27~\text{nm}$ can be practically realized, however, this period is quite attractive as it could provide the enhanced Dresselhaus spin-orbit coupling studied here along with a huge deterministic valley splitting \cite{McJunkin2021,Feng2022}.   

With regards to possible applications, the enhanced spin-orbit coupling of the wiggle well indicates that EDSR can be used for fast, electrically-driven manipulations of single-spin, Loss-DiVincenzo qubits without the use of micromagnets. Indeed, a fast, spin-orbit driven EDSR capability is one of the main attractive features of hole-spin qubits \cite{Maurand2016,Watzinger2018,Terrazos2021,Wang2022,Kloeffel2013}, and has recently also attracted interest in Si electron-spin qubits~\cite{Gilbert2023}. This possibility is supported by our calculations in Sec.~\ref{Results}, where it was shown that an EDSR Rabi frequency of $\Omega_{\text{Rabi}}/B > 500~\text{MHz/T}$ can be obtained near the optimal Ge oscillation wavelength $\lambda = 1.57~\text{nm}$. 
It is also possible that the enhanced spin-orbit coupling between the valleys may be used to drive fast singlet–triplet rotations near the valley-Zeeman hot spot \cite{Jock2022}. 
Finally, we mention that the enhanced and spatially varying spin-orbit coupling may have interesting effects on many-body physics in multi-electron dots \cite{Governale2002,Destefani2004,Corrigan2021,Ercan2021,Yannouleas2022}.

\section*{Acknowledgements}
Research was sponsored in part by the Army Research Office (ARO) under Award No. W911NF-17-1-0274 and No. W911NF-22-1-0090. 
The views, conclusions, and recommendations contained in this document are those of the authors and are not necessarily endorsed nor should they be interpreted as representing the official policies, either expressed or implied, of the Army Research Office (ARO) or the U.S. Government. The U.S. Government is authorized to reproduce and distribute reprints for Government purposes notwithstanding any copyright notation herein.

\appendix
\renewcommand\thefigure{\thesection.\arabic{figure}}    
\setcounter{figure}{0}

\section{Lattice constants in strained Si/SiGe quantum wells} \label{perpLatConstant}

To determine the lattice constants of the strained Si/SiGe heterostructure, we use pseudomorphic boundary conditions \cite{Van1986,Schaffler1997}, where the in-plane lattice constant $a_\parallel$ throughout the system is given by the relaxed lattice constant of the Si\textsubscript{0.7}Ge\textsubscript{0.3} barrier regions. This then also sets the lattice spacing along the growth direction as described below.


We take the relaxed lattice constant of a Si\textsubscript{1-$x$}Ge\textsubscript{$x$} alloy as
\begin{equation}
    a_o(x) = (1-x)a_{\text{Si}} + x a_{\text{Ge}},
\end{equation}
where $a_{\text{Si}} = 0.5431~\text{nm}$ and $a_{\text{Ge}} = 0.5657~\text{nm}$ are the relaxed lattice constants of Si and Ge, respectively.
Our structure therefore has an in-plane lattice constant $a_\parallel = a_o(0.3) = 0.5499~\text{nm}$ throughout the entire system. Atomic layers in our system with Ge concentration $n_{\text{Ge},n} \neq 0.3$ are therefore strained. Explicitly, the in-plane strain $\epsilon_{\parallel,n}$ of layer $n$ is
\begin{equation}
    \epsilon_{\parallel,n} = \frac{a_\parallel - a_o(n_{\text{Ge},n})}{a_o(n_{\text{Ge},n})}.
\end{equation} 
For a bulk Si\textsubscript{1-$x^\prime$}Ge\textsubscript{$x^\prime$} alloy under biaxial stress perpendicular to the $[001]$, we have strains $\epsilon_{xx} = \epsilon_{yy} = \epsilon_{\parallel}$ and $\epsilon_{zz} = \epsilon_{\perp}$ that are related by \cite{Van1986}
\begin{equation}
    \epsilon_{\perp}(x^\prime) = -2 \frac{C_{12}(x^\prime)}{C_{11}(x^\prime)} \epsilon_{\parallel},
\end{equation}
where $C_{12}$ and $C_{11}$ are elastic constants that depend on the Ge concentration $x^\prime$. For pure Si ($x = 0$), we have $C_{11}(0) = 165.8~\text{GPa}$ and $C_{12}(0) = 63.9~\text{GPa}$, while for pure Ge ($x = 1$), we have $C_{11}(1) = 131.8~\text{GPa}$ and $C_{12}(1) = 48.3~\text{GPa}$ \cite{Niquet2009}. For simplicity, we assume that the elastic constants vary linearly with the Ge concentration For a well region composed of Si\textsubscript{1-$x^\prime$}Ge\textsubscript{$x^\prime$}, we would then have the out-of-plane lattice constant,
\begin{equation}
    a_\perp(x^\prime) = \Big(1 + \epsilon_{\perp}(x^\prime)\Big)a_o(x^\prime).
\end{equation} In our system with its inhomogeneous Ge concentration profile, we take the lattice spacing between layers $n$ and $n+1$ with Ge concentrations $n_{\text{Ge},n}$ and $n_{\text{Ge},n+1}$, respectively, as $a_{\perp}^{(n+1,n)}/4$, where
\begin{equation}
    a_{\perp}^{(n+1,n)} = 
    \frac{1}{2}\Big(a_\perp(n_{\text{Ge},n+1}) + a_\perp(n_{\text{Ge},n})\Big),
\end{equation}
is the average of the out-of-plane lattice constants expected for strained regions with Ge concentration $n_{\text{Ge},n}$ and $n_{\text{Ge},n+1}$, respectively.

\section{Pseudospin basis transformation details} \label{PSBTM} In this appendix, we provide details of the pseudospin basis introduced in Sec. \ref{VCModel} of the main text. The pseudospin basis is defined as
\begin{equation}
    \ket{\boldsymbol{k}_\parallel n \bar{\nu} \bar{\sigma}} = \sum_{\nu \sigma} \ket{\boldsymbol{k}_\parallel n\nu\sigma} U^{(n)}_{\nu\sigma,\bar{\nu}\bar{\sigma}}, \label{PSdef}
\end{equation}
where $\bar{\nu}$ and $\bar{\sigma}$ label the new orbitals with $\bar{\sigma} = \Uparrow,\Downarrow$ being a \textit{pseudospin} label, $\nu$ and $\sigma$ are indices of the original basis with $\sigma = \uparrow,\downarrow$ simply denoting spin, and $U^{(n)}$ is the transformation matrix of layer $n$.

Our first requirement of the new basis is that it diagonalizes the onsite spin-orbit coupling. Following Chadi \cite{Chadi1977}, spin-orbit coupling is taken to be an intra-atomic (onsite) coupling between $p$ orbitals and enters into the Hamiltonian as the matrix $S$. The explicit form of the $S$ matrix of atom $j$ in atomic layer $n$ in the $p$-orbital subspace $\left\{\ket{p_z \uparrow},\ket{p_x \downarrow},\ket{p_y \downarrow},\ket{p_z \downarrow},\ket{p_x \uparrow},\ket{p_y \uparrow}\right\}$ is
\begin{equation}
     S^{(nj)} = \frac{\Delta^{(nj)}_{\text{SO}}}{3}
    \begin{pmatrix}
    0 & -1 & i & 0 & 0 & 0 \\
    -1 & 0 & i & 0 & 0 & 0 \\
    -i & -i & 0 & 0 & 0 & 0 \\
    0 & 0 & 0 & 0 & 1 & i \\
    0 & 0 & 0 & 1 & 0 & -i \\
    0 & 0 & 0 & -i & i & 0 \\
    \end{pmatrix}, \label{Smtx}
\end{equation}
where $\Delta_{\text{SO}}^{(nj)}$ is the spin-orbit energy. All matrix elements involving $s$, $s^{*}$, and $d$ orbitals are set to zero in $S$ and are not shown. Note that spin-orbit coupling, in principle, does exist between $d$-orbitals, but is much smaller than the $p$-orbital couplings and is typically neglected \cite{Jancu1998}. Also note that the spin-orbit energy depends on if the atom is Si or Ge, but the form of the spin-orbit coupling matrix is independent of atom type. It turns out that the $S$ matrix is diagonalized by the eigenstates of total angular momentum \cite{Winkler2003}. Within the $p$-orbital subspace, these states are given by
\begin{align}
    \left|p_1 \Uparrow\right> &= 
    \frac{i}{\sqrt{2}}\left(
    \left|p_x \downarrow\right> - i \left|p_y \downarrow\right>
    \right), \label{p1u}\\
    \left|p_2 \Uparrow\right> &= 
    \frac{1}{\sqrt{6}}\left(2 \left|p_z \uparrow\right>
    -\left|p_x \downarrow\right> - i \left|p_y \downarrow\right>
    \right), \\
    \left|p_3 \Uparrow\right> &= 
    \frac{1}{\sqrt{3}}\left(\left|p_z \uparrow\right>
    +\left|p_x \downarrow\right> + i \left|p_y \downarrow\right>
    \right), \\
    \left|p_1 \Downarrow\right> &= 
    \frac{i}{\sqrt{2}}\left(
    \left|p_x \uparrow\right> + i \left|p_y \uparrow\right>
    \right), \\
    \left|p_2 \Downarrow\right> &= 
    \frac{1}{\sqrt{6}}\left(2 \left|p_z \downarrow\right>
    +\left|p_x \uparrow\right> - i \left|p_y \uparrow\right>
    \right), \\
    \left|p_3 \Downarrow\right> &= 
    \frac{1}{\sqrt{3}}\left(\left|p_z \downarrow\right>
    -\left|p_x \uparrow\right> + i \left|p_y \uparrow\right>
    \right) \label{p3d},
\end{align}
where $\Uparrow,\Downarrow$ are \textit{pseudospin} labels. We find $\ket{p_1 \bar{\sigma}}$ and $\ket{p_2 \bar{\sigma}}$ both have an $S$ matrix eigenvalue of $\Delta^{(nj)}_{\text{SO}}/3$, while $\ket{p_3 \bar{\sigma}}$ has an eigenvalue of $-2\Delta^{(nj)}_{\text{SO}}/3$. We also define pseudospin $s$ and $s^{*}$ orbitals as
\begin{align}
    &\left|s \Uparrow\right> =
    \ket{s \uparrow}, \\
    &\left|s^* \Uparrow\right> =
    \ket{s^* \uparrow}, \\
    &\left|s \Downarrow\right> =
    \ket{s \downarrow}, \\
    &\left|s^* \Downarrow\right> =
    \ket{s^* \downarrow}, \label{ssd}
\end{align}
and pseudospin $d$-orbitals as
\begin{align}
    &\left|d_1 \Uparrow\right> = 
        \frac{i}{\sqrt{2}}\left(
        \left|d_{zx} \downarrow\right> - i \left|d_{yz} \downarrow\right>
        \right), \\
    &\left|d_2 \Uparrow\right> = 
    \frac{1}{\sqrt{6}}\left(2 \left|d_{z^2} \uparrow\right>
    -\left|d_{zx} \downarrow\right> - i \left|d_{yz} \downarrow\right>
    \right), \\
    &\left|d_3 \Uparrow\right> = 
    \frac{1}{\sqrt{3}}\left(\left|d_{z^2} \uparrow\right>
    +\left|d_{zx} \downarrow\right> + i \left|d_{yz} \downarrow\right>
    \right), \\
    &\left|d_4 \Uparrow\right> =
    \ket{d_{xy} \uparrow}, \\
    &\left|d_5 \Uparrow\right> =
    \ket{d_{x^2 - y^2} \uparrow}, 
\end{align}
\begin{align}
    &\left|d_1 \Downarrow\right> = 
    \frac{i}{\sqrt{2}}\left(
    \left|d_{zx} \uparrow\right> + i \left|d_{yz} \uparrow\right>
    \right), \\
    &\left|d_2 \Downarrow\right> = 
    \frac{1}{\sqrt{6}}\left(2 \left|d_{z^2} \downarrow\right>
    +\left|d_{zx} \uparrow\right> - i \left|d_{yz} \uparrow\right>
    \right), \\
    &\left|d_3 \Downarrow\right> = 
    \frac{1}{\sqrt{3}}\left(\left|d_{z^2} \downarrow\right>
    -\left|d_{zx} \uparrow\right> + i \left|d_{yz} \uparrow\right>
    \right) \\
    &\left|d_4 \Downarrow\right> =
    \ket{d_{xy} \downarrow}, \\
    &\left|d_5 \Downarrow\right> =
    \ket{d_{x^2 - y^2} \downarrow} \label{d5d}.
\end{align}
Note that all pseudospin $s$, $s^{*}$, and $d$-orbitals are trivially eigenstates of the spin-orbit matrix $S$ with eigenvalue $0$. 
In addition, note that for $j = 1,2,3$, $\ket{d_j \Uparrow}$ and $\ket{d_j \Downarrow}$ are found from $\ket{p_j \Uparrow}$ and $\ket{p_j \Downarrow}$, respectively, by letting $p_x, p_y, p_z \rightarrow d_{zx}, d_{yz}, d_{z^2}$. The other pseudospin $d$ orbitals and $s$ orbitals are trivially related to the original basis. Note that we adopt these altered $d$-orbital pseudospin states even though they possess no spin-orbit coupling such that we obtain the pseudospin structure of the hopping matrices in Eqs. (\ref{To} - \ref{TD}) of the main text. Failure to adopt these $d$-orbital pseudospin states would result in coupling between the pseudospin sectors for $\mathbf{k}_\parallel = 0$.

Secondly, we require the pseudospin basis to transform the Hamiltonian in such a way that the minimum unit cell (in the absence of an inhomogenous potential $V_n$) decreases from two sites to one site for $\mathbf{k}_\parallel = 0$. In other words, the Fourier-transformed hopping matrix, which is introduced in Eq. (\ref{Ttilde}) of the main text, must become site independent for $\mathbf{k}_\parallel = 0$. Naively adopting the orbitals defined in Eqs. (\ref{p1u} - \ref{d5d}) for every site does not fulfill this requirement. However, this requirement is fulfilled if we adopt the orbitals defined in Eqs. (\ref{p1u} - \ref{d5d}) if the $\ket{p_1 \Uparrow}$, $\ket{p_1 \Downarrow}$, $\ket{d_1 \Uparrow}$, $\ket{d_1 \Downarrow}$, $\ket{d_4 \Uparrow}$, and $\ket{d_4 \Downarrow}$, orbitals are multiplied by $(-1)^n$, where $n$ is the site index in the 1D chain. In other words, we flip the sign of these select orbitals on every other site. As stated in the main text, this alternating structure for the transformation matrix is due to the presence of two sublattices in the diamond crystal structure of Si as shown in Fig. \ref{FIG2} (a).

For clarity, we now provide the explicit form of $U^{(n)}$. We write $U^{(n)}$ as
\begin{equation}
U^{(n)} = 
    \begin{pmatrix}
     U_s^{(n)} & 0 & 0 \\
     0 & U_p^{(n)} & 0 \\
     0 & 0 & U_d^{(n)}
    \end{pmatrix},
\end{equation}
where $U_s^{(n)}$, $U_p^{(n)}$, and $U_d^{(n)}$ are matrix blocks which describe how the $s$, $p$, and $d$ orbitals transform, respectively. Here, $U_s^{(n)} = I_{4 \times 4}$ is just identity. The $p$ block is given by
\begin{equation}
    U_p^{(n)} = 
    \begin{pmatrix}
      0 & \frac{2}{\sqrt{6}} & \frac{1}{\sqrt{3}} & 0 & 0 & 0 \\
      0 & 0 & 0 & \frac{i(-1)^{n}}{\sqrt{2}} & \frac{1}{\sqrt{6}} & \frac{-1}{\sqrt{3}}  \\
      0 & 0 & 0 &  \frac{(-1)^{n+1}}{\sqrt{2}} & \frac{-i}{\sqrt{6}} & \frac{i}{\sqrt{3}} \\
      0 & 0 & 0 & 0 & \frac{2}{\sqrt{6}} & \frac{1}{\sqrt{3}} \\
      \frac{i (-1)^{n}}{\sqrt{2}} & \frac{-1}{\sqrt{6}} & \frac{1}{\sqrt{3}} & 0 & 0 & 0 \\
      \frac{(-1)^{n}}{\sqrt{2}} & \frac{-i}{\sqrt{6}} & \frac{i}{\sqrt{3}} & 0 & 0 & 0 \\
    \end{pmatrix},
\end{equation}
where the columns correspond to the pseudospin orbitals in the order $\{p_1 \Uparrow, p_2 \Uparrow, p_3 \Uparrow, p_1 \Downarrow, p_2 \Downarrow, p_3 \Downarrow\}$, and the rows correspond to the ``standard'' orbitals in the order $\{ p_z \uparrow, p_x \uparrow, p_y \uparrow, p_z \downarrow, p_x \downarrow, p_y \downarrow\}$. Finally, the $d$ block is 
\begin{widetext}
\begin{equation}
    U_d^{(n)} = 
    \begin{pmatrix}
     0 & \frac{2}{\sqrt{6}} & \frac{1}{\sqrt{3}} & 0 & 0 & 0 & 0 & 0 & 0 & 0 \\
     0 & 0 & 0 & 0 & 0 & \frac{i(-1)^{n}}{\sqrt{2}} & \frac{1}{\sqrt{6}} & \frac{-1}{\sqrt{3}}  & 0 & 0 \\
     0 & 0 & 0 & 0 & 0 & \frac{(-1)^{n+1}}{\sqrt{2}} & \frac{-i}{\sqrt{6}} & \frac{i}{\sqrt{3}} & 0 & 0 \\
     0 & 0 & 0 & (-1)^n & 0 & 0 & 0 & 0 & 0 & 0  \\
     0 & 0 & 0 & 0 & 1 & 0 & 0 & 0 & 0 & 0 \\
     0 & 0 & 0 & 0 & 0 & 0 & \frac{2}{\sqrt{6}} & \frac{1}{\sqrt{3}} & 0 & 0 \\
     \frac{i (-1)^{n}}{\sqrt{2}} & \frac{-1}{\sqrt{6}} & \frac{1}{\sqrt{3}} & 0 & 0 & 0 & 0 & 0 & 0 & 0 \\
     \frac{(-1)^{n}}{\sqrt{2}} & \frac{-i}{\sqrt{6}} & \frac{i}{\sqrt{3}} & 0 & 0 & 0 & 0 & 0 & 0 & 0 \\
     0 & 0 & 0 & 0 & 0 & 0 & 0 & 0 & (-1)^n & 0 \\
     0 & 0 & 0 & 0 & 0 & 0 & 0 & 0 & 0 & 1 \\
    \end{pmatrix},
\end{equation}
\end{widetext}
where the columns correspond to the pseudospin orbitals in the order $\{d_1 \Uparrow, d_2 \Uparrow, d_3 \Uparrow, d_4 \Uparrow, d_5 \Uparrow, d_1 \Downarrow, d_2 \Downarrow, d_3 \Downarrow, d_4 \Downarrow, d_5 \Downarrow\}$, and the rows correspond to the ``standard'' orbitals in the order $\{ d_{z^2} \uparrow, d_{zx} \uparrow, d_{yz} \uparrow, d_{xy} \uparrow, d_{x^2 - y^2} \uparrow, d_{z^2} \downarrow, d_{zx} \downarrow, d_{yz} \downarrow, d_{xy} \downarrow, d_{x^2 - y^2} \downarrow\}$. 
Notice the $(-1)^n$ factors that flips the signs of the $\ket{p_1 \Uparrow}$, $\ket{p_1 \Downarrow}$, $\ket{d_1 \Uparrow}$, $\ket{d_1 \Downarrow}$, $\ket{d_4 \Uparrow}$, and $\ket{d_4 \Downarrow}$ orbitals on every odd site. This allows for a unit cell containing only one site for $\mathbf{k}_\parallel = 0$ as described above.

\section{Symmetry argument for the hopping structure of the Rashba and Dresselhaus Hamiltonian components} \label{SymArg}

In Sec. \ref{ExpansionSec} of the main text, we found the Fourier transformed hopping matrix $\widetilde{T}^{(n)}_\pm(\mathbf{k}_\parallel)$ to have the form 
\begin{equation}
    \widetilde{T}^{(n)}_{\pm}(\mathbf{k}_\parallel) = \widetilde{T}^{(n)}_o +  \widetilde{T}^{(n)}_R(\mathbf{k}_\parallel) 
    \pm \widetilde{T}^{(n)}_D(\mathbf{k}_\parallel) + \mathcal{O}(\mathbf{k}_\parallel^2), \label{Texpansion2}
\end{equation}
where $\widetilde{T}^{(n)}_R(\mathbf{k}_\parallel)$ and $\widetilde{T}^{(n)}_D(\mathbf{k}_\parallel)$ are the Rashba and Dresselhaus hopping matrices, respectively, and are given by
\begin{align}
    \widetilde{T}^{(n)}_R(\mathbf{k}_\parallel)
    &=  \Phi^{(n)} 
    \left(k_y \sigma_x - k_x \sigma_y\right), \label{TRashba2}\\
    \widetilde{T}^{(n)}_D(\mathbf{k}_\parallel)
    &= \Phi^{(n)}\left(k_x\sigma_x - k_y \sigma_y\right), \label{TD2}
\end{align}
with $\Phi^{(n)}$ being a real-valued $10 \times 10$ matrix with vanishing diagonal elemental. Importantly, the sign in front of the Rashba hopping matrix $\widetilde{T}^{(n)}_R$ is site independent, while the sign of the Dresselhaus hopping matrix $\widetilde{T}^{(n)}_D$ changes sign between every site, as indicated by the $\pm$ in Eq. (\ref{Texpansion2}).
This can be understood as originating from the diamond crystal structure of the Si by the following symmetry argument; Let us consider the case of pure Si such that $\left\{\Phi^{(n)},\widetilde{T}^{(n)}_R,\widetilde{T}^{(n)}_D\right\} \rightarrow \left\{\Phi,\widetilde{T}_R,\widetilde{T}_D\right\}$ all lose their dependence on the layer index. Next, note that under a $C_4$ rotation about the $z$-axis (growth axis), we have $\{k_x, k_y, \sigma_x, \sigma_y\} \rightarrow \{k_y, -k_x, \sigma_y, -\sigma_x\}$. This leaves invariant the Rashba term in Eq. (\ref{TRashba2}) and flips the sign of the Dresselhaus term in Eq. (\ref{TD2}). Finally, performing the same $C_4$ rotation on our diamond crystal structure in Fig. \ref{FIG2} (a) transforms red atoms into blue atoms and vice versa in the sense that the nearest neighbor vectors of even and odd atomic layers swap. In other words, the two sublattices of the Si lattice swap. This in turn, swaps $\widetilde{T}_+$ and $\widetilde{T}_-$ within the tight binding chain.  
Clearly then, the Rashba hopping term $\widetilde{T}_{R}$, being invariant under a $C_4$ rotation, should contain the part common to $\widetilde{T}_+$ and $\widetilde{T}_-$.  In contrast, the Dresselhaus hopping term $\widetilde{T}_D$ should contain the part which is different between $\widetilde{T}_+$ and $\widetilde{T}_-$, since it flips sign under a $C_4$ rotation. This then explains the $\pm$ in front of $\widetilde{T}^{(n)}_D$ in Eq. (\ref{Texpansion2}).

\section{$\Omega$ and $\Phi$ matrices} \label{OmegaPhi}

In Sec. \ref{ExpansionSec} of the main text, we introduced the $10 \times 10$ matrices $\Omega^{(n)}$ and $\Phi^{(n)}$ as components of the hopping matrices in Eqs. (\ref{To} - \ref{TD}). These matrices can be further decomposed into the block forms,
\begin{align}
    \Omega^{(n)} &= 
    \begin{bmatrix}
    \Omega^{(n)}_{00} & \Omega^{(n)}_{01} & 0 \\
    -\Omega_{01}^{(n)T} & \Omega^{(n)}_{11} & 0  \\
    0 & 0 & \Omega^{(n)}_{22}
    \end{bmatrix}, \\
    \Phi^{(n)} &= 
    \begin{bmatrix}
    \Phi_{00}^{(n)} & \Phi_{01}^{(n)} &  \Phi_{02}^{(n)} \\
    \Phi_{01}^{(n)T} & \Phi_{11}^{(n)} & \Phi_{12}^{(n)} \\
    -\Phi_{02}^{(n) T} & \Phi_{12}^{(n) T} & 0
    \end{bmatrix},
\end{align}
where the shapes of the diagonal blocks are $5 \times 5$, $4 \times 4$, and $1 \times 1$, respectively, and the diagonal block matrices satisfy $\Omega_{ii}^{(n)T} = \Omega_{ii}^{(n)}$ and $\Phi_{ii}^{(n)T} = -\Phi_{ii}^{(n)}$. Note that this implies that all diagonal elements of $\Phi^{(n)}$ are zero. Here, the orbital ordering used is $\left\{s, s^*, p_1, d_2, d_3, p_2, p_3, d_1, d_4, d_5\right\}$. Generically, these matrices depend on the layer index $n$ due to Ge concentration changing from layer to layer. In the case of a uniform Ge concentration, however, this layer dependence goes away, $\left\{\Omega^{(n)}, \Phi^{(n)}\right\} \rightarrow \left\{\Omega, \Phi\right\}$. In the particular case of an unstrained Si system, the $\Omega$ matrix blocks (in eV) are given by
\begin{align}
    \Omega_{00} &= 
    \begin{pmatrix}
    -3.73 & -2.78 & 0 & 0 & 0 \\
    -2.78 & -9.03 & 0 & 0 & 0 \\
    0 & 0 & -0.73 & -1.71 & 2.42 \\
    0 & 0 & -1.71 & 1.63 & 2.03 \\
    0 & 0 & 2.42 & 2.03 & 0.19
    \end{pmatrix}, \\
     \Omega_{11} &= 
    \begin{pmatrix}
    0.73 & 0 & 1.71 & -2.42  \\
    0 & 0.73 & -2.42 & -1.71  \\
    1.71 & -2.42 & 1.24 & 0  \\
    -2.42 & -1.71 & 0 & 1.24  \\
    \end{pmatrix}, \\
     \Omega_{22} &= 
    3.06, \\
     \Omega_{01} &= 
    \begin{pmatrix}
    2.74 & 1.94 & 0 & -2.59  \\
    2.89 & 2.05 & 0 & -0.90  \\
    2.15 & -3.04 & 0.19 & 0 \\
    -2.07 & -1.60 & 0.70 & -2.92 \\
    -1.60 & 0.94 & -0.99 & -2.07
    \end{pmatrix},
\end{align}
and the $\Phi$ matrix blocks (in eV $\cdot$ \AA) are given by 
\begin{align}
     \Phi_{00} &= 
    \begin{pmatrix}
    0 & 0 & 3.23 & 1.43 & -2.03 \\
    0 & 0 & 3.40 & 0.50 & -0.70 \\
    -3.23 & -3.40 & 0 & -1.25 & -0.89 \\
    -1.43 & -0.50 & 1.25 & 0 & 1.72 \\
    2.03 & 0.70 & 0.89 & -1.72 & 0
    \end{pmatrix}, \\
     \Phi_{11} &= 
    \begin{pmatrix}
    0 & 3.57 & -0.15 & -0.10  \\
    -3.57 & 0 & -0.1 & 0.15  \\
    -0.15 & -0.1 & 0 & -1.16  \\
    -0.10 & -0.15 & 1.16 & 0  \\
    \end{pmatrix}, \\
     \Phi_{02} &= 
    \begin{pmatrix}
    0 & 0 & 2.66 & 1.72 & 2.43
    \end{pmatrix}, \\
     \Phi_{12} &= 
    \begin{pmatrix}
    -1.54 & 2.17 & 2.98 & 0
    \end{pmatrix}.
\end{align}
Note that these $\Omega$ and $\Phi$ are precisely the matrices used in the simplified model of Sec. \ref{Mechanism}, where Ge atoms are treated as Si atoms with orbitals shifted up by a constant energy $E_\text{Ge}$.

%

\end{document}